\documentclass[twocolumn]{aastex631}

\usepackage{color}
\usepackage{amsmath}
\usepackage{natbib}

\hypersetup{breaklinks=true, colorlinks=true, urlcolor={blue}, citecolor={blue}, linkcolor={blue}}

\accepted{Jan.~18, 2022}
\shortauthors{Hu et al.}

\newcommand{\LD}[1]{{\color{black} #1}}

\newcommand{\chandra}{\emph{Chandra}}

\newcommand{\lumcgs}{erg~s$^{-1}$}

\begin{document}
\title{A Comprehensive Study of the Spectral Variation and the Brightness Profile of Young Pulsar Wind Nebulae}
\author[0000-0001-8551-2002]{Chin-Ping Hu}
\affiliation{Department of Physics, National Changhua University of Education, Changhua, 50007, Taiwan}
\author[0000-0002-7005-7139]{Wataru Ishizaki}
\affiliation{Yukawa Institute for Theoretical Physics, Kyoto University, Kitashirakawa-Oiwake-cho, Sakyo-ku, Kyoto 606-8502, Japan}
\author[0000-0002-5847-2612 ]{C.-Y. Ng}
\affiliation{Department of Physics, The University of Hong Kong, Pokfulam Road, Hong Kong}
\author[0000-0002-8796-1992]{Shuta J. Tanaka}
\affiliation{Department of Physical Sciences, Aoyama Gakuin University, Kanagawa, Sagamihara, 252-5258, Japan}
\author{Y.-L. Mong}
\affiliation{School of Physics and Astronomy, Monash University, Clayton, VIC 3800, Australia}

\correspondingauthor{C.-P. Hu}
\email{cphu0821@gm.ncue.edu.tw}

\begin{abstract}
We present a systematic study of particle transport by diffusion in young pulsar wind nebulae (PWNe). We selected nine bright sources that are well resolved with the \emph{Chandra X-ray Observatory}. We analyzed archival data to obtain their radial profiles of photon index ($\Gamma$) and surface brightness ($\Sigma$) in a consistent way. These profiles were then fit with a pure diffusion model that was tested on Crab, 3C 58 and G21.5$-$0.9 before. In addition to the spectral softening due to the diffusion, we calculated the synchrotron power and built up the theoretical surface brightness profile. For each source, we performed separate fits to the $\Gamma$ and the $\Sigma$ profiles. We found these two profiles of most PWNe are similar, except for Crab and Vela. \LD{Both} profiles can be well described by our model, suggesting that diffusion dominates the particle transport in most sampled PWNe.  The discrepancy of parameters between the $\Gamma$ profile and $\Sigma$ profiles is relatively large for 3C 58 and G54.1+0.3. This difference could be attributed to the elongated shape, reflecting boundary, and the non-uniform magnetic field. Finally, we found no significant correlations between diffusion parameters and physical parameters of PWN and pulsar. 
\end{abstract}

\section{Introduction}\label{introduction}
An isolated pulsar releases most of its rotational energy through high-energy relativistic particle outflow called pulsar wind. 
When the pulsar wind interacts with the surrounding medium and is over-pressured, a pulsar wind nebula (PWN) is formed. This drives a wind-termination shock, at which the particles are heated and re-accelerated to ultra-relativistic regime \citep[see reviews by][]{GaenslerS2006, Slane2017}. The shocked particles move through the magnetic field in a PWN. They cool down and emit synchrotron radiation from radio to X-rays, and inverse-Compton radiation in $\gamma$-rays.

Observationally, the injected particle distribution could be inferred from the broadband spectral-energy distribution of PWNe. A one-zone model without spatial information can reproduce the observed properties \citep[see, e.g.][]{PaciniS1973, TanakaT2010, BucciantiniAA2011, VorsterTF2013}. The improvement of spatial resolution of instruments makes it possible to investigate the detailed structure and the evolution of PWNe. When transporting outward, the high-energy particles lose their energies on a fast timescale and result in a softening of the X-ray spectrum. At the same time, the synchrotron flux decreases. The radial profiles in the radio band have been first modeled as a diffusion nebula with a particle source injected at the center \citep{Gratton1972, Wilson1972}. A one-dimensional magnetohydrodynamic (MHD) model with a time-independent steady flow has been established from this idea and tested on Crab \citep{KennelC1984a, KennelC1984b, PorthKK2014, PorthVL2016}. 

However, this model is not applicable for a few other PWNe under the unprecedented spatial resolution of \emph{Chandra X-ray Observatory} \citep[see e.g.,][]{SlaneHS2004}. Combining the photon index ($\Gamma$) profile and the surface brightness ($\Sigma$) profile, \citet{IshizakiTA2017} found that the simple MHD model with steady stream fails to reproduce observed properties in 3C 58 and G21.5$-$0.9, indicating that spatial diffusion of particles, which has been modeled by \citet{TangC2012}, should be taken into account. The pure diffusion model was first proposed to explain the flux and the spectral index of Crab in the optical band \citep{Wilson1972}. The morphology of a young PWN is approximated by a sphere that results from the diffusion of an input particle source into the surrounding space.  Later X-ray observations on 3C 58 and G21.5$-$0.9 show similar $\Gamma$ distribution and suggest the importance of diffusion in these PWNe \citep{BocchinoSC2005, SlaneCS2000, SlaneHS2004, GuestST2019}. Moreover, \citet{PorthVL2016} and \citet{IshizakiAK2018} conclude that the particle transfer in these two objects is dominated by diffusion through modeling the surface brightness. In addition, the $\gamma$-ray halo observed in Geminga also indicates that diffusion plays an important role in PWNe \citep{LindenAB2017, AbeysekaraAA2017, BaoFB2021}.

Although these works suggest the importance of diffusion, a systematic test of the pure diffusion model is lacking. We aim to test whether this model can be applied to the $\Gamma$ and $\Sigma$ profiles in young PWNe that are observed with \chandra. The sample selection is described in Section \ref{sample_selection}, in which nine PWNe are selected.  In Section \ref{spectral_analysis}, we describe the X-ray data reduction and the spectral analysis. We perform joint fit of multiple \chandra\ observations and calculate the observational $\Gamma$ and $\Sigma$ profiles. The detailed theoretical modeling is described in Section \ref{diffusion_model}. We follow the model derived by \citet{TangC2012} with transmitting  boundary\LD{,i.e., particles are not reflected by the boundary and the emission from particles beyond the boundary is not taken into account, and} use the full synchrotron spectrum to calculate the $\Gamma$ and the $\Sigma$ profiles \citep{IshizakiTA2017, IshizakiAK2018}. The fitting results are shown in Section \ref{result} and discussed in Section \ref{discussion}. Finally, we summarize our work in Section \ref{summary}.

\begin{deluxetable*}{llcccccc}
\tablecaption{Physical properties and observational parameters for the young PWNe in this study \label{tab:pwn_list}} 
\tablehead{\colhead{PWN} & \colhead {PSR} & \colhead {Distance} & \colhead{$\tau_{\rm{PWN}}^a$} & \colhead {$B_{\rm{PWN}}^a$} & \colhead {log $L_{\rm{PWN}}$} & \colhead {log $L_{\rm{sd}}$} & \colhead{Exp.}\\
\colhead{} & \colhead{} & \colhead{(kpc)} & \colhead{(kyr)} & \colhead{($10^{-5}$ G)} & \colhead{(erg s$^{-1}$)} & \colhead{(erg s$^{-1}$)} & \colhead{(ks)}  }
\startdata
3C 58 & J0205+6449 & 2.0 & 2.5 & 1.7 & 31.1 & 37.4 & 391.9 \\
G21.5$-$0.9 & J1833$-$1034 & 4.4 & 1.0 & 4.7 & 33.1 & 37.75 & 749.0\\
G54.1+0.3 & J1930+1852 & 4.9 & 1.7 & 1.0 & 32.0 & 37.1 & 325.6\\
G291.0$-$0.1$^b$ & unknown & 3.5 & 1.2 & 7.0 & 30.5 & 37.7 & 472.6\\
G310.6$-$1.6 & J1400$-$6325 & 7.0 & 0.92 & 1.7 & 33.0 & 37.7 & 231.5\\
Kes 75 & J1846$-$0258 & 5.8 & 0.7 & 2.0 & 33.3 & 36.9 & 383.1 \\
MSH 15$-$5\textit{2} & B1509$-$58 & 5.2 & 1.7 & 1.5 & 32.5 & 37.5 & 244.2 \\
Vela & B0833$-$45 & 0.287 & 6.5 & 0.49 & 30.3 & 37.5  & 630.1 \\
Crab & B0531+21 & 2.0 & 0.95 & 8.5 & 34.6 & 38.7  & 23.4
\enddata
\tablenotetext{a}{These two quantities are adopted from \citet{TanakaT2011, TanakaT2013}.}
\tablenotetext{b}{An X-ray source CXOU J111148.6$-$603926 is suggested to be the pulsar powers G291.0$-$0.1 although the timing properties remains unknown. The spin-down power is estimated from the broadband spectral fit of the PWN \citep{SlaneHT2012}.}
\end{deluxetable*}

\section{Sample Selection}\label{sample_selection}
Our PWN sample consists of young and bright sources obtained from Table 1 in \citet{KargaltsevP2010} that are well resolved by \chandra\ with angular size over 20\arcsec.  Nine PWNe, including 3C 58, G21.5$-$0.9, G54.1+0.3, G291.0$-$0.1, G310.6$-$1.6, Kes 75, MSH 15$-$5\textit{2}, Vela, and Crab, are selected in this study. Their basic properties are listed in Table \ref{tab:pwn_list}. The distances, $B$-field strengths, PWN radius, and age are adopted from literatures \citep[see e.g.,][and references therein]{TanakaT2011, TanakaT2013}. We obtain all the pulsar properties from the ATNF pulsar catalog\footnote{\href{https://www.atnf.csiro.au/research/pulsar/psrcat/}{https://www.atnf.csiro.au/research/pulsar/psrcat/}} \citep{Manchester2005}.

We analyzed the \chandra\ observations taken with the Advanced CCD Imaging Spectrometer (ACIS) and operated in the timed-exposure (TE) mode for each PWN. For G21.5$-$0.9, we excluded those observations for calculation purposes, with the focal plane offset or the CCD not at optimal temperature. 
Moreover, we only analyze recent \chandra\ observations (after 2012) of Crab to reduce the computing time. All the data sets used in this analysis are summarized in Table \ref{tab:chandra_log}. The radial profile of Crab, 3C 58, and G21.5$-$0.9 have been investigated before \citep[see, e.g.,][]{SlaneHS2004, BambaAD2010, TangC2012, GuestST2019}. We, however, re-analyzed the data on our own to ensure the latest calibration is applied and consistent with other PWNe in our sample. 

Note that G10.65+2.96 fits our selection criterion. However, we cannot obtain the difference in spectral properties of inner and outer regions. \LD{G11.2$-$0.3 is one of the youngest PWN \citep{BorkowskiRR2016}. It has an elongated shape and most of the PWN emission overlapped the jet viewing from Earth. After removing the jet \citep[see, e.g., Figure 5 in][]{BorkowskiRR2016}, the remaining photons are not statistically significant to measure the spectral difference between the inner and outer PWN. } We excluded them from our samples. 

The central pulsar of G291.0$-$0.1 remains controversial. A young pulsar J1105$-$6107 was suggested as the central pulsar of this PWN \citep{KaspiBM1997}. However, a later broadband study suggested \LD{that J1105$-$6107 is well outside the PWN and unlikely related \citep{SlaneHT2012}.  A spin-down luminosity of $\sim 5.7\times 10^{37}$ \lumcgs\ of the central object CXOU J111148.6$-$603926 is suggested although this is  estimated from the broadband spectral fit to the PWN \citep{SlaneHT2012}. We note that this is not a precise measurement since a large uncertainty is possible. } 


\section{Spectral Analysis}\label{spectral_analysis}

All the data reduction and basic analysis were carried out using the Chandra Interactive Analysis of Observations (CIAO) version 4.11 with the calibration database of version 4.8.3. We reprocessed the data using the task \texttt{chandra\_repro}.  We use \texttt{reproject\_obs} to reproject all data sets of a PWN to create a merged event file, and use \texttt{flux\_obs} to create an exposure-corrected image. These images were used to detect the position of pulsars and other point sources. The innermost circular region within 2.5\arcsec\ radius of the pulsar was removed in the subsequent spectral analysis. We then extracted the PWN spectra from circular annulus regions centered on the pulsar. We adjusted the width of annuli to obtain a sufficient signal-to-noise ratio in each bin (Figure \ref{fig:pwn_image}). Several targets have jets\LD{, which are known to have different spectral behaviors compared to other parts of PWNe and may have different mechanisms of particle transfer \citep[see, e.g.,][]{SlaneHS2004, TeminSR2010}. Moreover, a few objects like MSH 15$-$5\textit{2} have knots or blobs at the opposite direction of jets. We followed previous analysis to remove those structures as well as background point sources from our spectral analysis \citep[see, e.g.,][]{YatsuKS2009}. }

The source and background spectra, corresponding weighted response matrix files (RMFs), and ancillary response files (ARFs) of each annulus region are extracted using \texttt{specextract} command. The background is selected from a nearby source-free region on the same chip. Following the standard analysis procedure for extended sources, we generated weighted ARFs by setting \texttt{weight=yes} and apply no point-source aperture correction by setting \texttt{correctpsf=no}.

We performed all the spectral analysis using the \texttt{Sherpa} environment. The energy range of the spectral fitting is restricted to 0.5--8 keV to optimize the signal-to-noise ratio except for 3C 58, where we fit its high-energy tail in the 1--8 keV range to avoid the contribution of a soft thermal component below $\sim1$~keV \citep{SlaneHS2004, GotthelfHN2007}.  All the spectra are grouped to have at least 30 X-ray photons in each energy bin. We use an absorbed power-law (PL) model to fit the spectra jointly across all observations for each PWN. We chose absorption model \texttt{tbabs} and set the interstellar abundance according to \citet{WilmsAM2000}, in which uses the cross-section presented in \citet{VernerYB1993}. 

We first fit the entire PWN with a single absorbed PL model to constrain the hydrogen column density $N_{\rm{H}}$. This is justified if we assume that the $N_{\rm{H}}$ is generally the same among the entire PWN and the spectral behavior of each PWN is steady. Then, we fixed $N_{\rm{H}}$ at the best-fit value and fit the spectra of each annulus by linking $\Gamma$ and the normalization across all observations. The purpose is to reduce the computation time. A few time-variable softening/hardening areas of some PWNe have been reported, but they do not significantly change the overall spectrum of each annulus \citep[see, e.g.,][]{GuestS2020}. We calculated the 0.5--8 keV unabsorbed flux using \texttt{sample\_flux} function with $10^4$ \LD{iterations} to obtain the $\Sigma$ profile of each PWN. \LD{For each iteration, this command simulates parameter values from a normal distribution around best-fit parameters. }

\section{Pure Diffusion Model}\label{diffusion_model}
A model based on diffusion was proposed to improve the MHD model, in which the cross-field scattering is small and the particle diffusion can be ignored \citep{deJagerD2009}. In the pure diffusion model, the diffusion coefficient ($D$) is energy-independent. We further assume that $D$ and the $B$-field strength ($B_{\rm{PWN}}$) are constants in the PWN and set the outer boundary to be transparent for particles. \LD{Previous studies have suggested that additional advection of particles could improve the fit, but this only dominates the innermost region of Crab \citep{TangC2012, IshizakiTA2017}}. On the other hand, diffusion dominates most of the volume of the PWN. 

\begin{figure*}[ht]
\begin{minipage}{0.3\linewidth}
\includegraphics[width=0.95\textwidth]{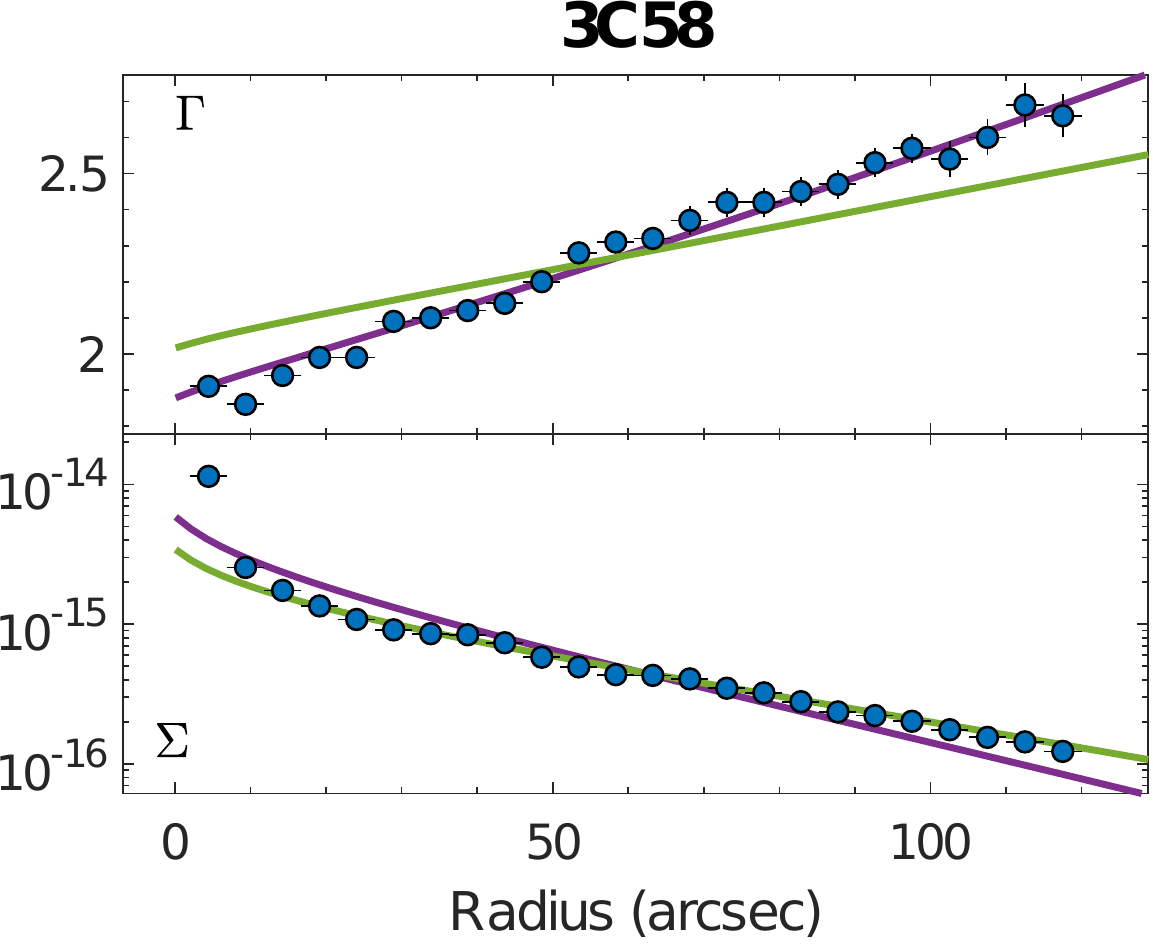}\\
\includegraphics[width=0.95\textwidth]{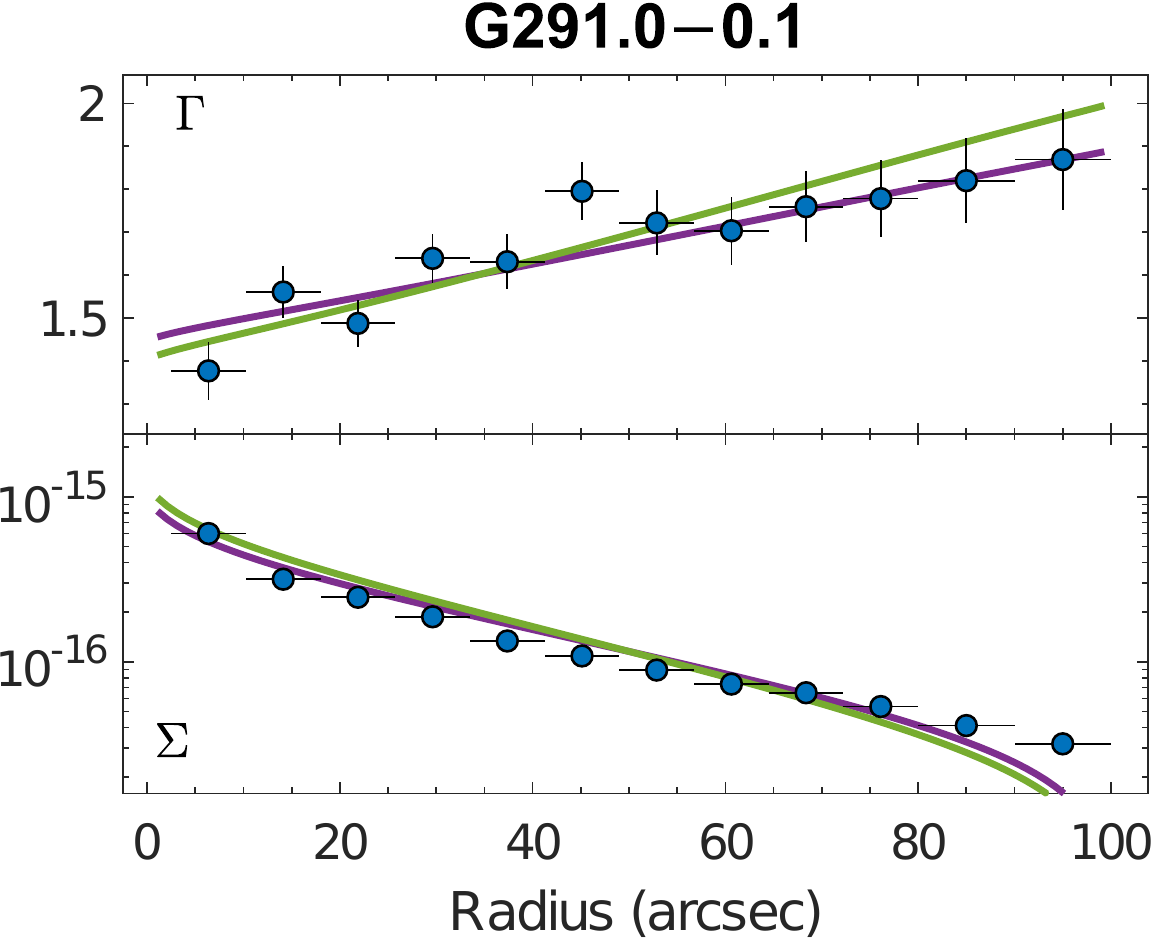}\\
\includegraphics[width=0.95\textwidth]{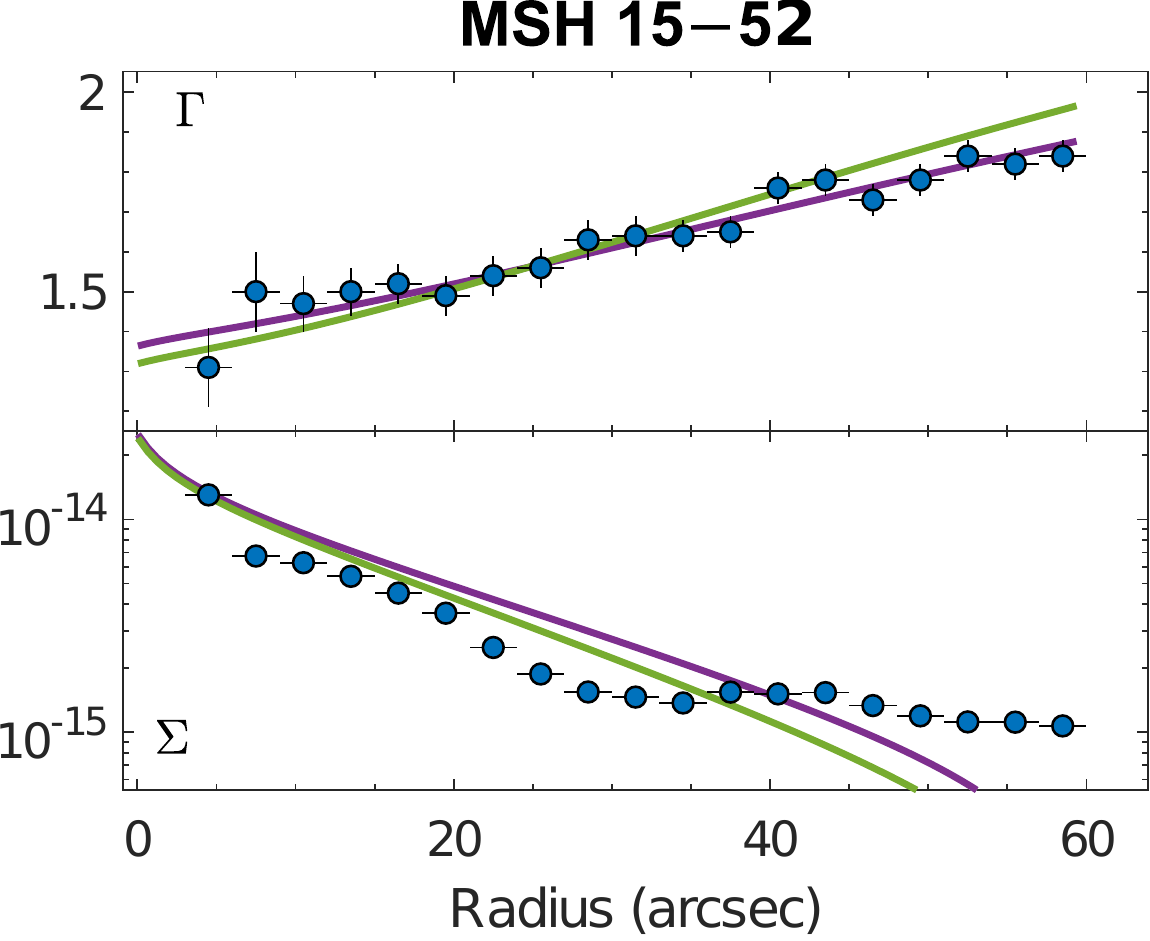} 
\end{minipage}
\smallskip
\begin{minipage}{0.3\linewidth}
\includegraphics[width=0.95\textwidth]{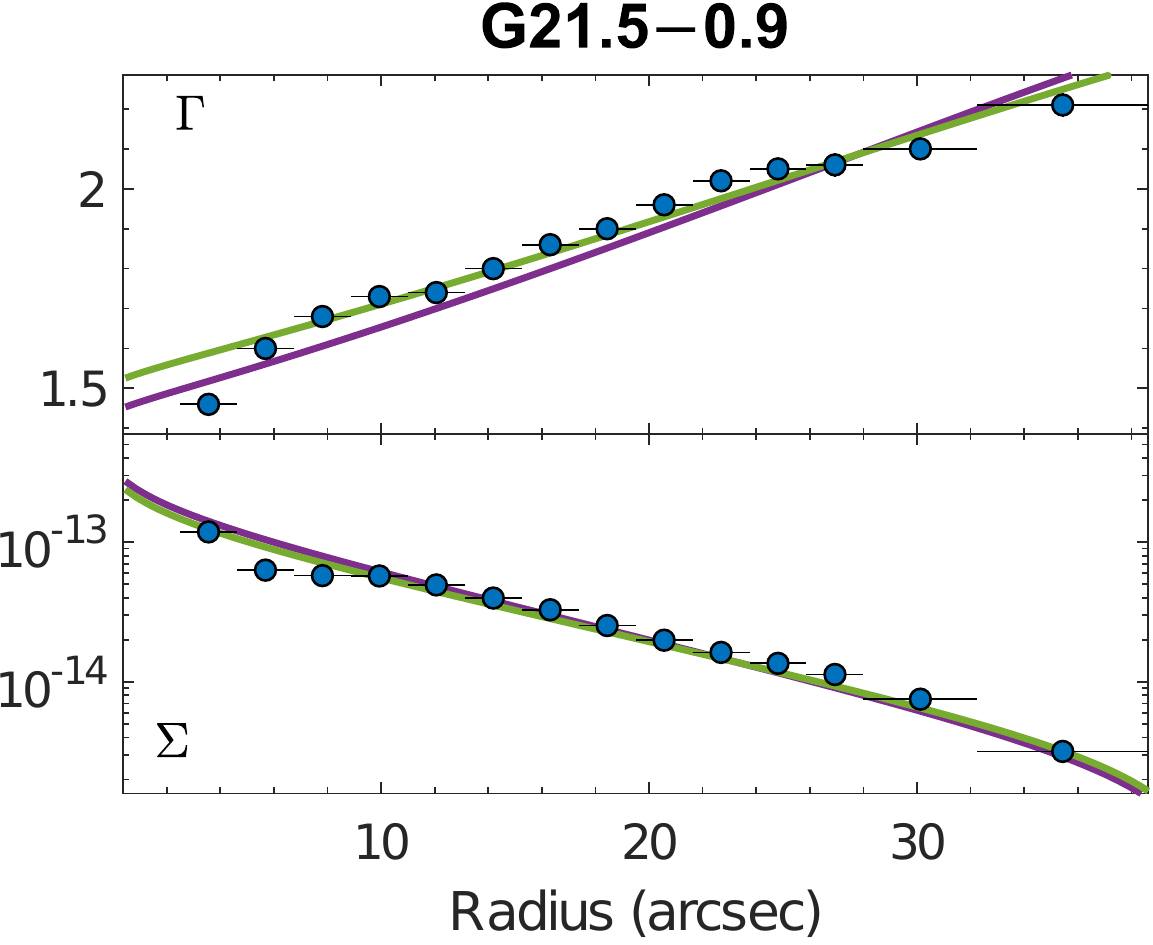} \\[0.2cm]
\includegraphics[width=0.95\textwidth]{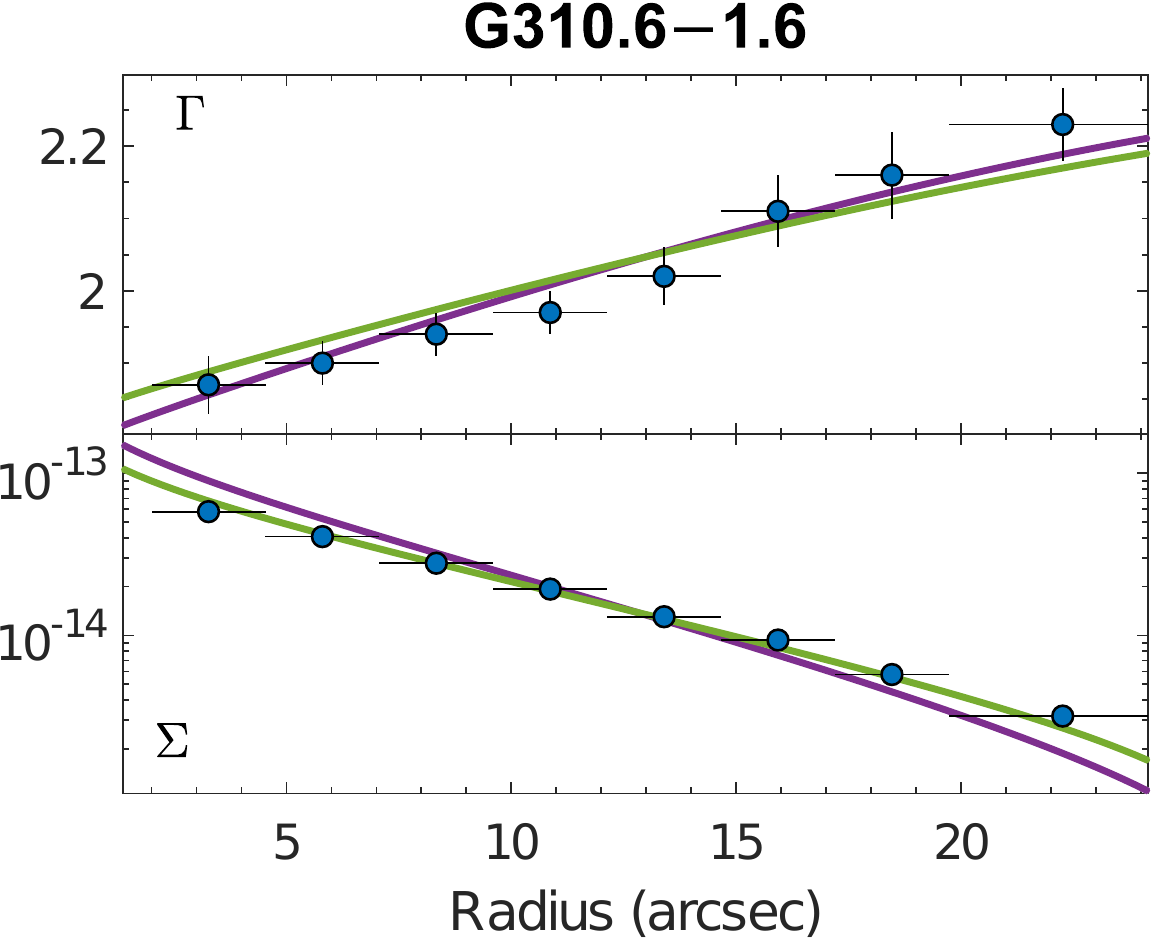} \\[0.2cm]
\includegraphics[width=0.95\textwidth]{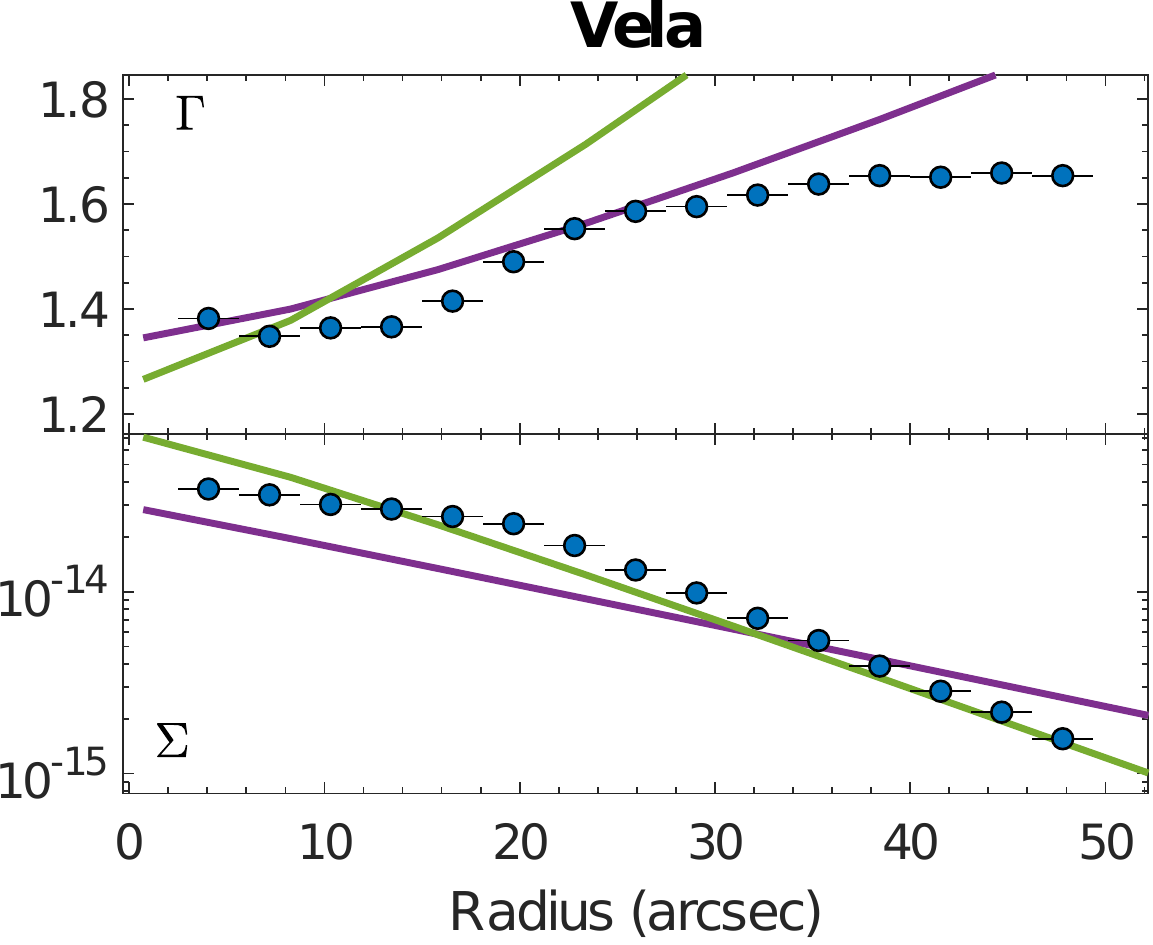} 
\end{minipage}
\smallskip
\begin{minipage}{0.3\linewidth}
\includegraphics[width=0.95\textwidth]{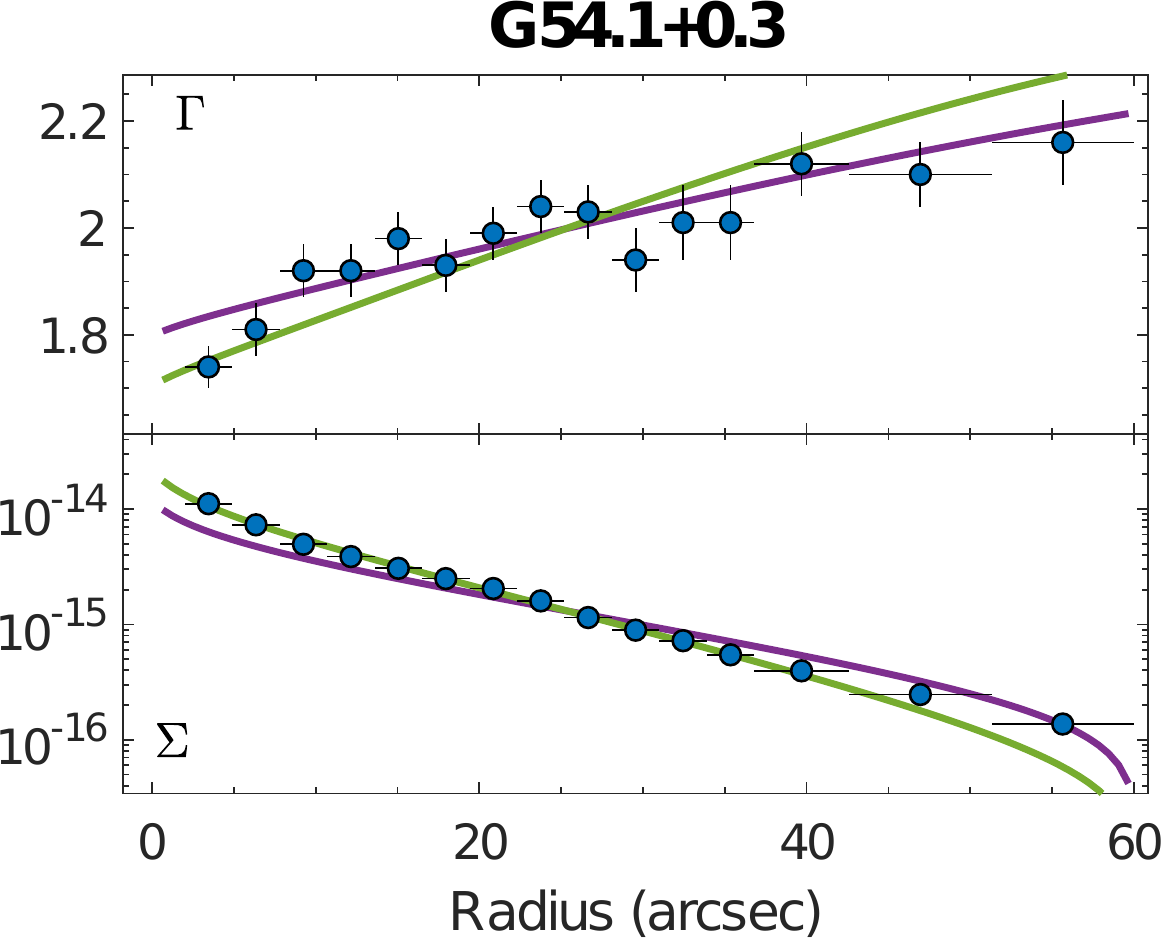} \\[0.2cm]
\includegraphics[width=0.95\textwidth]{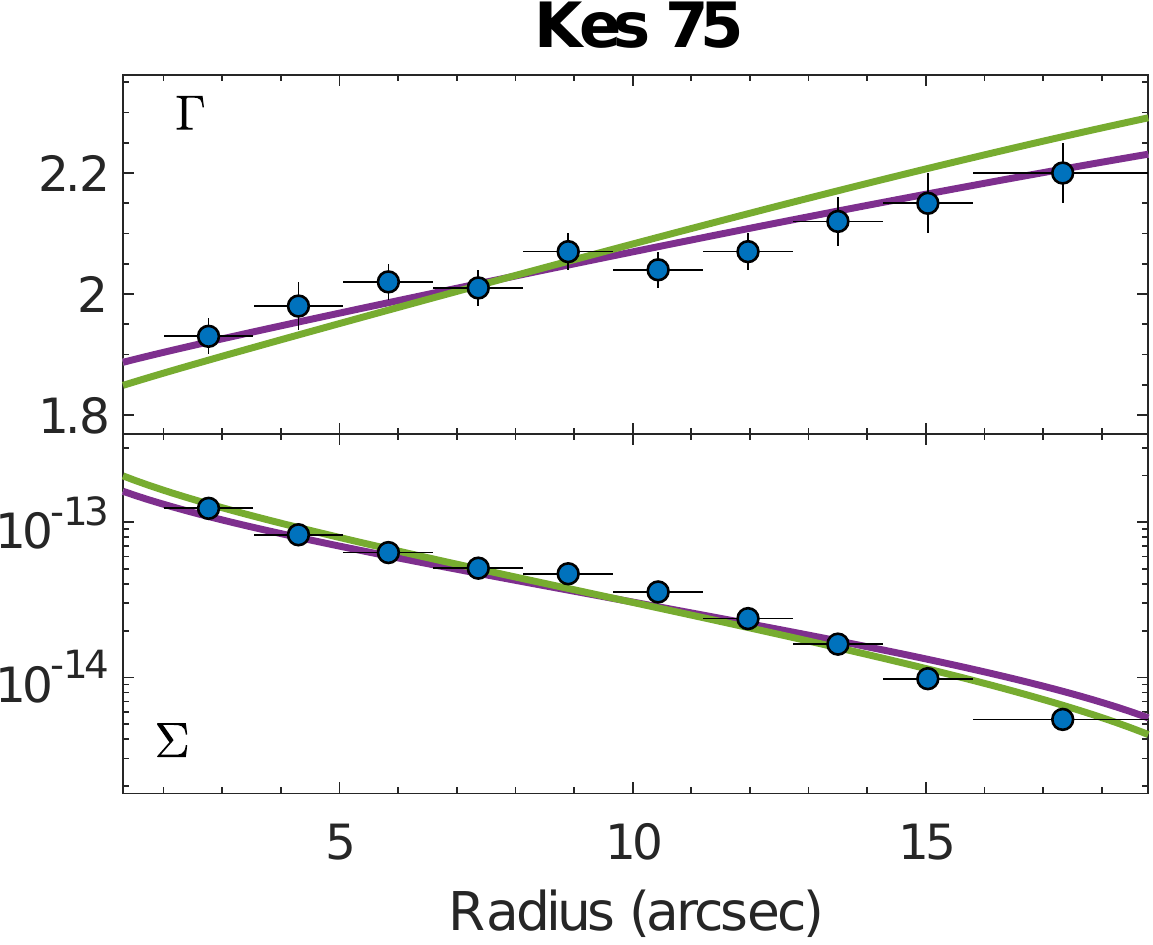} \\[0.2cm]
\includegraphics[width=0.95\textwidth]{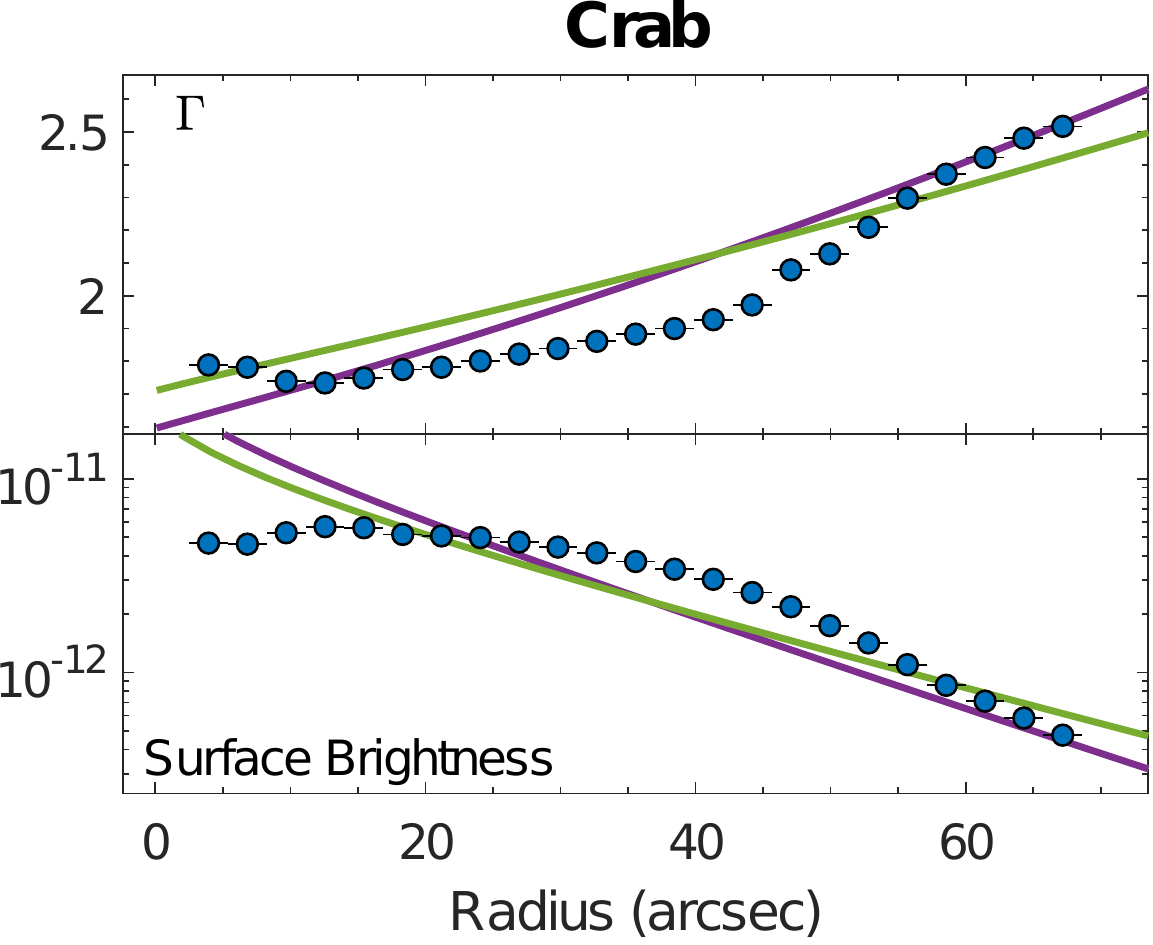}
\end{minipage}
\caption{The radial profile of $\Gamma$ and the unabsorbed surface brightness of for our PWN sample.For each PWN, the upper panel shows the the radial profiles of $\Gamma$, and the lower panel shows the $\Sigma$ profile in units of erg cm$^{-2}$ s$^{-1}$ arcsec$^{-2}$. The purple curve denotes the best \LD{$\Gamma$-fit for} the $\Gamma$ profile. The green curve denotes the best \LD{$\Sigma$-fit} that fits the $\Sigma$ profile. \label{fig:pwn_image_result} }
\end{figure*}

We assume power-law distributed particles $\dot{N}(E, r=0)=KE^{-\alpha}$ injected from the center of a PWN, where $E$ is the particle energy and $\alpha$ is the power-law index. The particles diffuse from the center and distribute spherically symmetry. As the particles extend to a radius $R$, which corresponds to the outer bound of a PWN, we adopt a simple case that the particles transmit out and leave the system. The $B$-field strength and the PWN radius $R$ are pre-determined parameters using the broadband spectral-energy distribution fitting results obtained from \citet{TanakaT2011, TanakaT2013}. (see Table \ref{tab:pwn_list}).  Following \citet{Gratton1972}, the particle number density distribution is
\begin{equation}\label{eq:number_density}
    N(E,r,t)=\frac{K}{4\pi rD}E^{-\alpha}f_{\alpha}(u,v),
\end{equation}
where $f_{\alpha}(u,v)$ is an integral
\begin{equation}\label{eq:integration_f}
    f_{\alpha}(u,v)=\frac{1}{\sqrt{\pi}}\int_v^{\infty} \left( 1-\frac{u}{x} \right)^{\alpha-2}\frac{e^{-x}}{\sqrt{x}}dx,
\end{equation}
and $u$ is defined as
\[
u=\frac{r^2QE}{4D},
\]
where
\[
Q=\frac{4\sigma_T}{3m^2c^3}\frac{B^2}{8\pi}
\]
is the factor of synchrotron energy loss.
The lower limit of the integral in Equation \ref{eq:integration_f} is 
\[
v=\begin{cases}
\displaystyle \frac{r^{2}}{4Dt} & \displaystyle t<\frac{1}{QE}\\[0.3cm]
\displaystyle \frac{r^{2}QE}{4D} & \displaystyle t>\frac{1}{QE}
\end{cases}
\]
where $t$ is the age of the PWN. Once we obtain the energy distribution of particles, we calculate the synchrotron spectrum at any radius with an assumed magnetic field. The normalization constant $K$ can be derived from the luminosity of the PWN. We assume a fraction $\eta$ of the spin-down energy that is converted to the energy of particles above 1 TeV. This can be written as
\begin{equation}
    \int_{E_0}^{E_{\rm{max}}} E\dot{N}(E, r=0, t) dE=\eta L_{\rm{sd}},
\end{equation}
where $E_0=1~{\rm TeV}$ is the lower bound of the integration, and $E_{\rm{max}}$ is the maximum value of the electrostatic energy in the polar cap of the central pulsar. By performing an integration to $\dot{N}(E, r=0, t)=KE^{-\alpha}$, we obtain
\begin{equation}
K=\frac{\eta L_{\rm sd}\left(\alpha-2\right)}{1-\left(E_{\rm max}/E_0\right)^{2-\alpha}}E_0^{\alpha-2}.
\end{equation}

\begin{figure}[ht]
\centering
\includegraphics[width=0.49\textwidth]{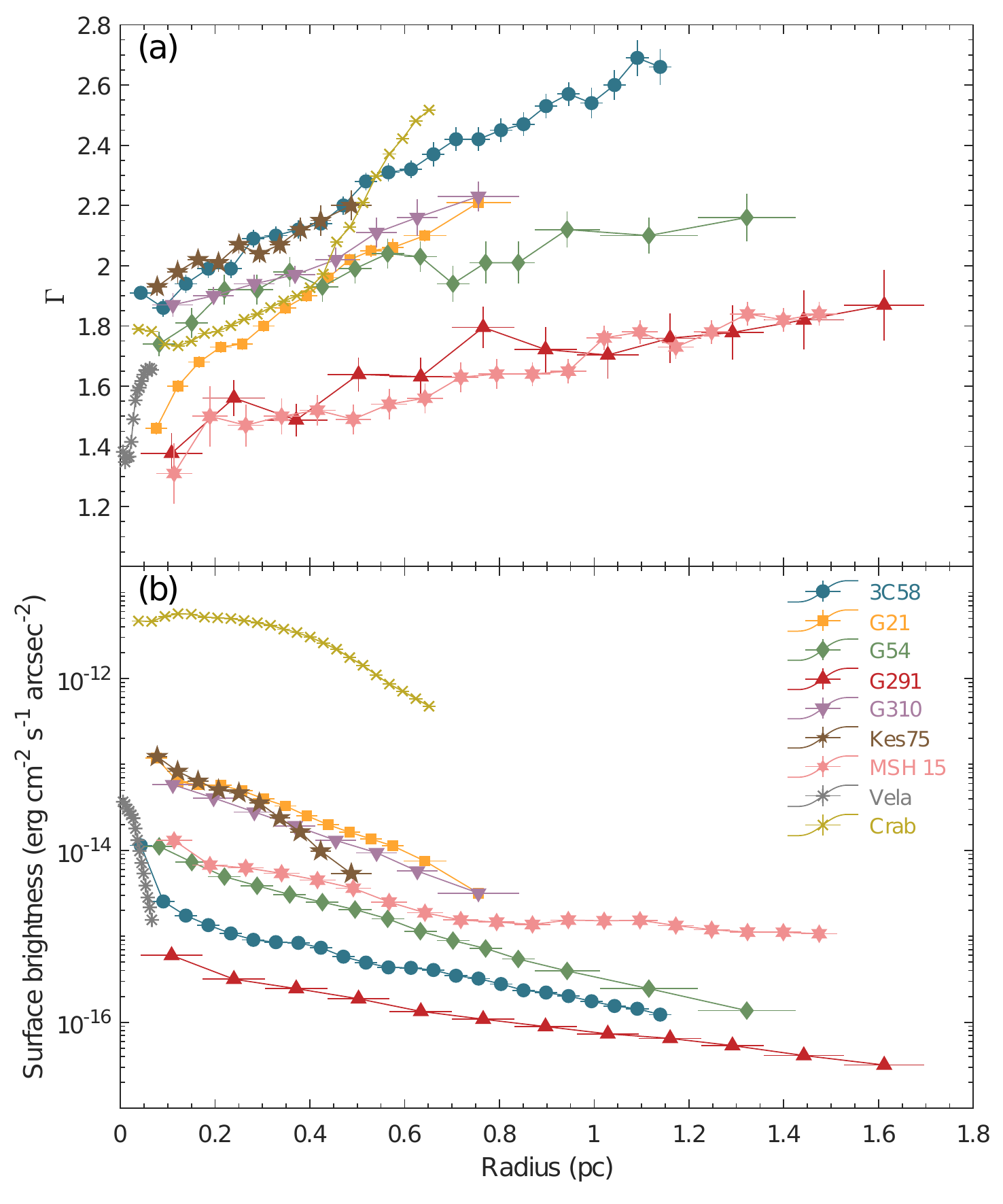} 
\caption{$\Gamma$ and unabsorbed $\Sigma$ profiles of all PWNe. The x-axis is the physical size in units of parsec. The size is converted from the angular size and assumed distances listed in Table \ref{tab:pwn_list}.   \label{fig:pwn_result_all}}
\end{figure}

In our analysis, we use the full synchrotron spectrum to normalize the energy distribution of particles for the emission calculation. We followed the procedure presented in \citet{TanakaT2010, TanakaT2011} and \citet{IshizakiTA2017} to calculate the synchrotron power. The spectral flux density $j_{\nu}(r)$ can be written as
\begin{equation}
j_\nu (r)=\int N(E,r,t)p_{\nu}(E)dE,
\end{equation}
where
\begin{equation}
p_\nu(E)=\frac{\sqrt{3}e^3B}{mc^2}F\left(\frac{\nu}{\nu_c}\right),
\end{equation}
\begin{equation}
\nu_c=\frac{3}{4\pi}\frac{eB}{mc}\gamma^2,
\end{equation}
$F(x)=x\int_x^\infty K_{5/3}(y)dy$ with $x=\nu/\nu_c$, and $K_{5/3}$ is the modified Bessel function of order $5/3$ \citep{RybickiL1986}.  The \LD{specific} surface brightness $M_\nu$ can be integrated as
\begin{equation}
M_\nu(r_\perp)=\int_C\frac{j_\nu(r)}{4\pi}ds,
\end{equation}
where $ds$ is a line element along the line-of-sight $C$, $r_\perp$ is the distance perpendicular to the line of sight from the central pulsar, i.e. angular distance. In a spherically symmetric system, this line integral can be written explicitly as
\begin{equation}
M_\nu(r_\perp)=2\int_{r_\perp}^{r_N}\frac{j_\nu(r)}{4\pi}\frac{rdr}{\sqrt{r^2-r_\perp^2}},
\end{equation}
where $r_N$ is a radius of the nebula.
\LD{The surface brightness $\Sigma(r_\perp)$ can be obtained by integrating over the energy range from $\nu_1$ to $\nu_2$ as follows:}
\begin{equation}
\Sigma(r_\perp)=\int_{\nu_1}^{\nu_2}M_{\nu}(r_\perp)d\nu.
\end{equation}
Finally, the spectral index $\Gamma$ (from $\nu_1$ to $\nu_2$) can be calculated as
\begin{equation}
    \Gamma=\frac{\log[M_{\nu_1}(r_\perp)/M_{\nu_2}(r_\perp)]}{\log(\nu_1/\nu_2)}.
\end{equation}

\begin{deluxetable*}{llcccc}
\tablecaption{Fitting Result of Selected PWNe \label{tab:pwn_fit_result}} 
\tablehead{\colhead{PWN Name} & \colhead {Model$^a$} & \colhead {$\alpha$} & \colhead {$D$ } & \colhead{$\eta$} & \colhead{$DB^{-3/2}$} \\
 & & & \colhead{($10^{25}$ cm$^2$ s$^{-1}$)} & \colhead{($\times10^{-3}$)} & \colhead{($10^{32}$ cm$^2$ s$^{-1}$ G$^{-3/2}$)}  } 
\startdata
3C 58 & $\Gamma$ & 2.6 & $12$ & 12.3 & $17$  \\
 & $\Sigma$ & 2.9 & $30$ & 37.8 & $43$ \\
G21.5$-$0.9 & $\Gamma$ & 1.75 & $16$ & 56.1 & $4.9$ \\
 & $\Sigma$ & 1.9 & $20$ & 50 & $6.2$ \\
G54.1+0.3 & $\Gamma$ & 2.5 & $11$ & 66.8 & $35$  \\
 & $\Sigma$ & 2.3 & $45$ & 35.3 & $14$ \\
G291.0$-$0.1 & $\Gamma$ & 1.7 & $220$ & \LD{0.6} & $38$\\
 & $\Sigma$ & 1.8 & $350$ & \LD{0.8} & $60$\\
G310.6$-$1.6 & $\Gamma$ & 2.46 & $5$ & 68.5 & $7.1$\\
 & $\Sigma$ & 2.55 & $7.8$ & 84.7 & $11$ \\
Kes 75 & $\Gamma$ & 2.6 & $6.5$ & 537 & $7.3$ \\
 & $\Sigma$ & 2.5 & $4$ & 362 & $4.4$ \\
MSH 15$-$5\textit{2} & $\Gamma$ & \LD{1.6} & \LD{$19$} & \LD{52.6} & \LD{$33$} \\
 & $\Sigma$ & \LD{1.5} & \LD{$13$} & \LD{40} & \LD{$22$}\\
Vela &  P & 1.4 & $0.025$ & 0.62 & $0.23$ \\
 & $\Sigma$ & 1.2 & $0.0085$ & 0.59 & $0.078$ \\
Crab & $\Gamma$ & 2.01 & $30$ & 340.2 & $3.8$ \\
 & $\Sigma$ & 2.25 & $50$ & 349.8 & $6.4$
\enddata
\tablenotetext{a}{The $\Gamma$ model fits the $\Gamma$ profile and the $\Sigma$ model fits the $\Sigma$ profile. }
\end{deluxetable*}

\section{Fitting results} \label{result}
We fit the radial profiles of $\Gamma$ and $\Sigma$ with the pure diffusion model described in Section \ref{diffusion_model}. \LD{Three parameters, $\alpha$, $D$, and $\eta$ are tuned to fit the profiles, while others parameters are fixed.} \LD{Our objective is to determine if the simple diffusion could reproduce the characteristic of the profile. Therefore, we did not perform a formal $\chi^2$-fit because} it is difficult to discuss the goodness of fit statistically owing to the large systematic uncertainty of this simple model. The best-fit results are listed in Table \ref{tab:pwn_fit_result}. 
We performed two fits for each source: one fits the $\Gamma$ profile (\LD{$\Gamma$ fit}) and another fits the $\Sigma$ profile (\LD{$\Sigma$ fit}). Both profiles for each PWN are shown in Figure \ref{fig:pwn_image_result}. \LD{Note that these two fits are not independent. 
The shape of both profiles are determined once $\alpha$ and $D$ are chosen. 
The parameter $\eta$ only tunes the normalization of the $\Sigma$ profile. 
We discuss the parameter dependence in Section \ref{discussion}. }

The pure diffusion model generally well describes the sample PWNe. Both \LD{$\Gamma$- and $\Sigma$-fits} yield consistent results in G21.5$-$0.9, G291.0$-$0.1, G310.6$-$1.6, and Kes 75. These PWNe have no complex structure compared to Crab, Vela, and MSH 15$-$5\emph{2}. However, some PWNe are relatively faint with large uncertainties in $\Gamma$ that could smear detailed structures in the profile. On the other hand, large discrepancies between the parameters yielded from the \LD{$\Gamma$- and $\Sigma$-fits} are seen in 3C 58, G54.1+0.3, and possibly MSH 15$-$5\textit{2}.  There are however a couple of poor fits, in particular Crab and Vela. They have significant changes in slope in both profiles, and their brightness profiles are concave down. \LD{Moreover, these are especially observed in the innermost region, which the advection may play an important role \citep{TangC2012, IshizakiTA2017}.}

In Figure \ref{fig:pwn_result_all} we plot the $\Gamma$ and $\Sigma$ profiles versus the physical size. All the PWNe have a similar shape in both profiles except for Crab and Vela. The $\Gamma$ monotonically and smoothly increases with distance from the pulsar in nearly all sources.  The overall slope resides in a narrow range. All the sampled PWNe have similar $\Sigma$ profiles except for Crab and Vela. We discuss these apparent features and the physical interpretation using the diffusion model in Section \ref{discussion}.

\section{Discussion}\label{discussion}

\subsection{Applicability of the Pure Diffusion Model}

Our result shows that the pure diffusion model can explain the radial profiles of $\Gamma$ and $\Sigma$ in soft X-rays for most young PWNe, except for a few cases. We suggest the pure diffusion model could be applied to other young PWNe in general. Except for Crab,  none of the sampled PWNe show an abrupt change in $\Gamma$ profile, suggesting that the steady flow assumed in the MHD model cannot describe the observed phenomena \citep{SlaneHS2004, TangC2012, IshizakiTA2017}. A few PWNe have large discrepancies between the \LD{$\Gamma$- and $\Sigma$-fits} parameters, implying further improvements to the model may be needed. Moreover, the profiles of Crab and Vela cannot be described with the diffusion model. Here we discuss the discrepancy between the $\Gamma$ and the $\Sigma$ profiles and the detailed features of individual PWNe.

\LD{The spin-down power of the central pulsar of G291 remains unclear. With the assumption of $L_{sd}=5.7\times10^{37}$~erg~s$^-1$, both the $\Gamma$ and $\Sigma$ profiles can be fit with $\alpha\sim1.7$ and $\eta\lesssim0.001$. This suggests a low energy conversion fraction comparable to that of Vela. Similar low ratio of the X-ray luminosity and the spin-down luminosity was also suggested by \citet{SlaneHT2012}. This is low but remains within a reasonable range of known pulsars. We also tried to use different values of $L_{sd}$ in the range between $\sim10^{36}$~erg~s$^-1$ and $5.7\times10^{37}$~erg~s$^-1$. All values gives the same fitting with the same parameters except for $\eta$.}

The discrepancy between the \LD{$\Gamma$- and $\Sigma$-fits} is significant in 3C 58 and G54.1+0.3. The best-fit $D$ of \LD{$\Gamma$ and $\Sigma$ fits} differ from each other at a factor larger than 2. We note that these sources are more elongated than others, such that the $B$-field could be different in the axial and equatorial directions. It was suggested that the viewing geometry of 3C 58 is close to edge-on, while G54.1+0.3 has an inclination angle close to $\approx147^{\circ}$ between the torus axis and the line of sight \citep{NgR2004, NgR2008}.  The non-spherical diffusion and the viewing geometry could result in the discrepancy between the $\Gamma$ and the $\Sigma$ profiles. This hints at the limitation of our simple spherical symmetric PWN under the pure diffusion approximation for the synchrotron emissivity calculation. \LD{Moreover, the age of 3C 58 remains controversial, as its connection to SN 1181 is not clear \citep{Kothes2013, RitterPL2021}. If 3C 58 is not as young as other sampled PWNe, the interaction between the PWN and the supernova remnant has to be taken into account, especially for the outer region of the PWN.} Finally, the energy-dependence of $D$, the reflection effect of the PWN boundary, the advection flow, and the inhomogeneity of the $B$ field may be further considered to improve the fit \citep{TangC2012}. 

Crab and Vela are the only two cases that clearly cannot be described by our diffusion model. It was argued that the advection is significant in Crab \citep{TangC2012}. If this is true, the magnetic field is expected to change spatially.  On the other hand, Vela shows an opposite trend as Crab in the $\Gamma$ profile, suggesting a different mechanism.  Considering that the distance to Vela is at least one order of magnitude closer than other PWNe in our sample, the observed emission region could only represent the very central part of the entire PWN \citep[see, e.g.,][]{KargaltsevP2008}. Therefore, it may not be easy to interpret it in the same way as other sources. Moreover, Vela is old comparing with the other PWNe. \LD{The interaction between the supernova remnant (SNR) and the PWN has to be taken into account. However, the jet and torus structure is not distorted by the reverse shock \citep[see, e.g.,][]{NgR2004}. A more plausible possibility is that the advection dominates the particle transfer in the innermost region of a PWN.} In addition, the spatial structure of the magnetic field could be another possibility to make it difficult to describe the profile with our simple model. Similar to Crab and Vela, MSH 15$-$5\textit{2} shows clear inner-ring and torus structures. A valley is seen in the brightness profile at 20--40\arcsec\ (0.5--1\,pc), which was reported by \citet{YatsuKS2009}. Therefore, a simple spherically symmetric or constant $B$-field approximation may not be justified. 
\begin{figure}[ht]
\centering
\includegraphics[width=0.49\textwidth]{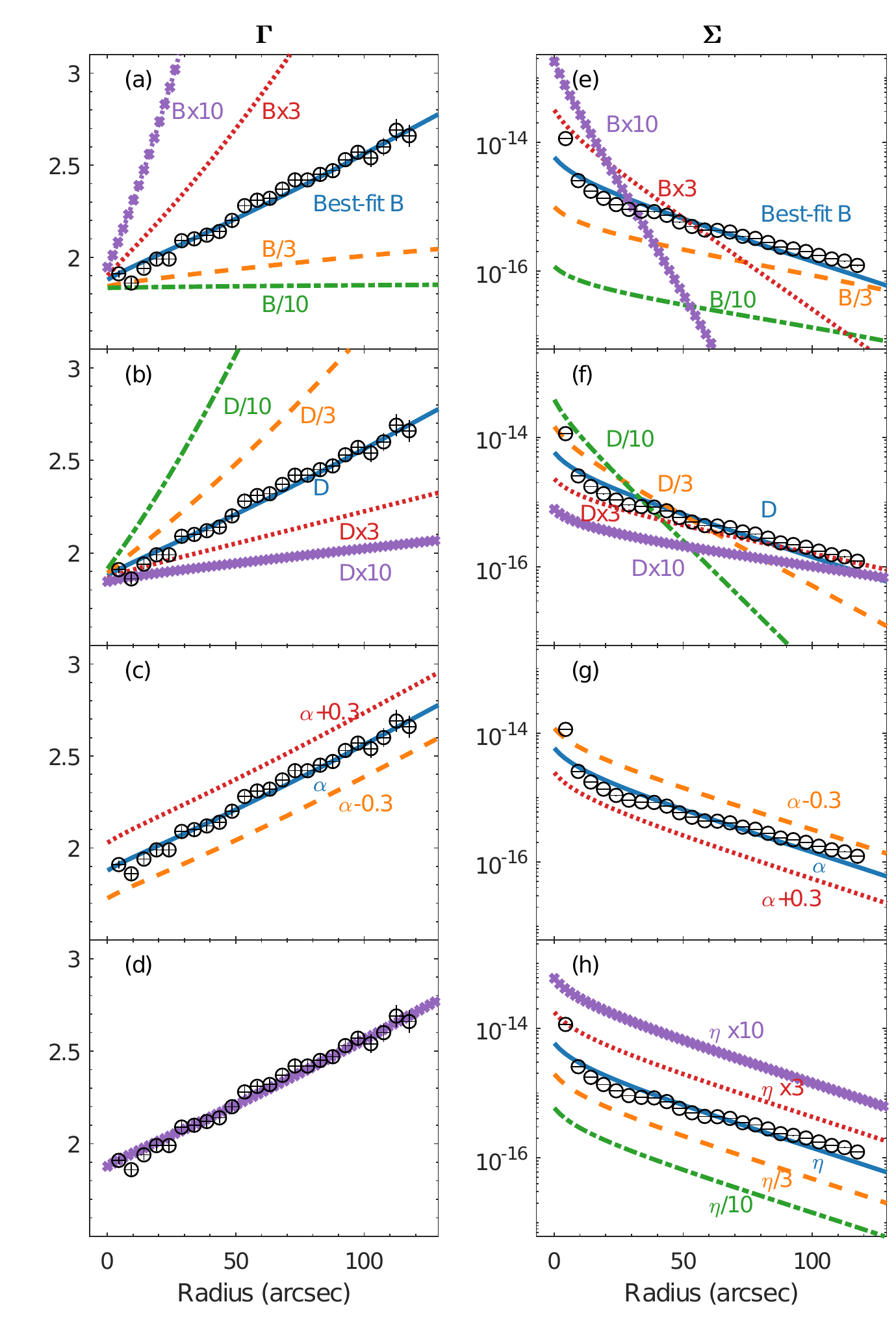} 
\caption{Parameter dependence of model profiles in $\Gamma$ (a)--(d) and $\Sigma$ (e)--(h) in units of erg cm$^{-2}$ s$^{-1}$ arcsec$^{-2}$. Data points are results from 3C 58.  \label{fig:parameter_degeneracy}}
\end{figure}

\begin{figure*}
\centering
\includegraphics[width=0.7\textwidth]{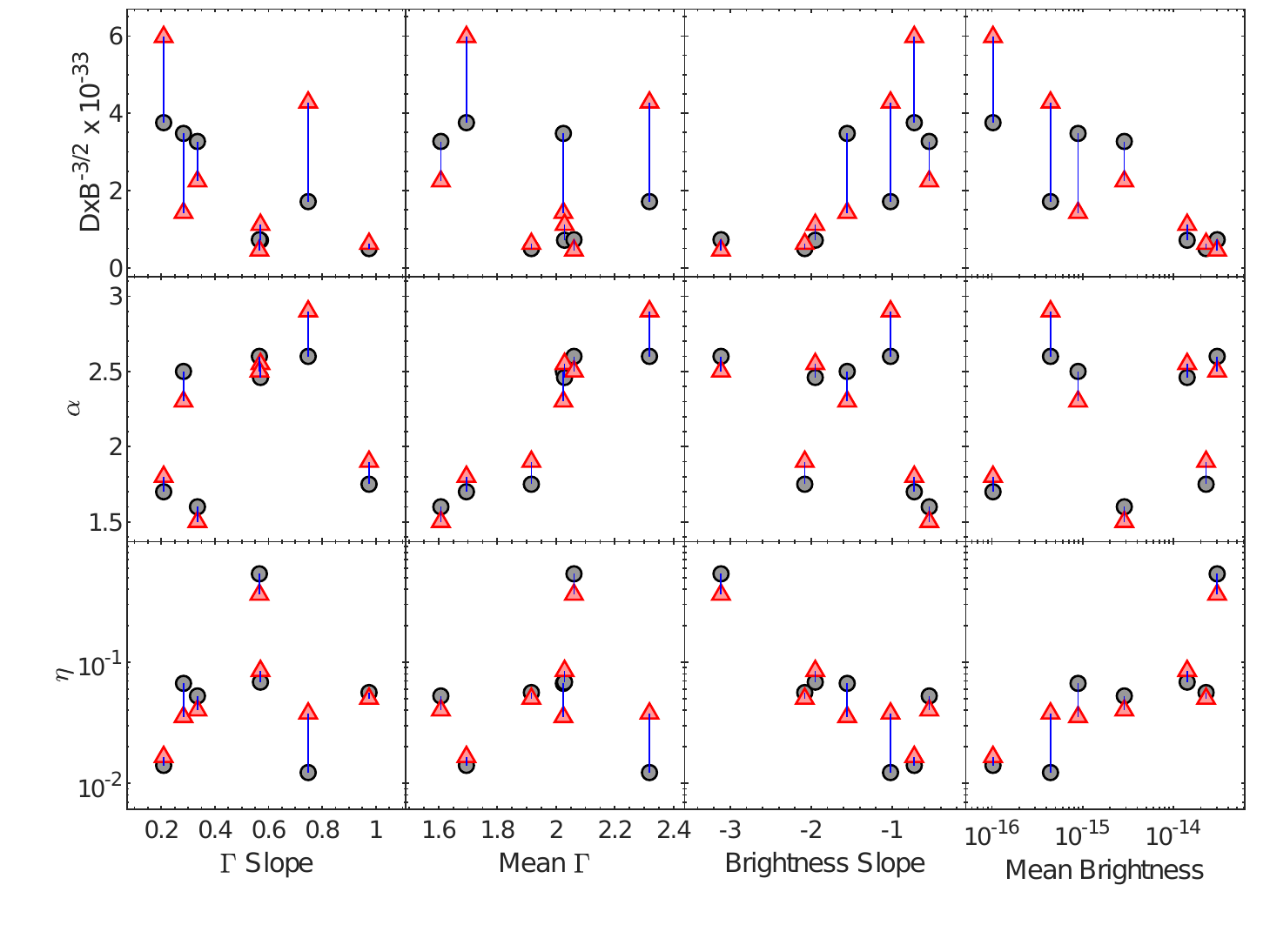} 
\caption{Correlation between model parameters and apparent parameters of the $\Gamma$ and $\Sigma$ profiles. Gray circles denote the parameters derived from the $\Gamma$ fit, while the red triangles are parameters from the $\Sigma$ fit. We use blue line to link the parameters obtained by \LD{$\Gamma$ and $\Sigma$ fits} for each PWN.  \label{fig:apparent_vs_model}}
\end{figure*}

\subsection{Degeneracy of Model Parameters}\label{parameter_degen}
Based on our pure diffusion model with transmitting  boundary, three parameters are critical for the $\Gamma$ profile: $B$ field in the PWN, $D$, and the injected power-law index $\alpha$ of the particle energy. Moreover, $\eta$ plays an important role in the $\Sigma$ profile. \LD{However, the $B$ field strength is a pre-defined value and varies dramatically among studies, see, e.g., G21 \citep{TanakaT2011, TanakaT2013, GuestST2019}. This parameter is coupled with $D$ and therefore it is meaningless to directly compare $D$ among PWNe.} Taking 3C 58 as an example, we demonstrate how the model profile changes as we change these parameters in Figure \ref{fig:parameter_degeneracy}. From Figures \ref{fig:parameter_degeneracy} (a) and (b), the effect of $B$ and $D$ are coupled with each other \citep{TangC2012}. A flatter $\Gamma$ profile, which implies a lower energy dissipation, can be achieved by both decreasing $B$ and increasing $D$, and vice versa. This is because the spatial extent of the X-ray emitting particles is defined by the distance over which they spread before exhausting their energy via synchrotron radiation. The ratio of the diffusion length of a particle emitting a photon of a given frequency $\nu$ to a radius $R$ can be written as $\sqrt{4Dt_{\rm cool}}/R\propto \left(R^{-2}\nu^{-1/2}DB^{-3/2}\right)^{1/2}$, where $t_{\rm cool}=1/QE$ is the cooling time scale of the synchrotron emission. The $\Sigma$ profile is the same if $DB^{-3/2}$ remains constant. Therefore, $DB^{-3/2}$ is the key parameter for the shape of the $\Gamma$ profile. Since the PWN $B$-field strength is difficult to estimate and it often depends on many assumptions, we will not discuss the effect of $D$. Instead, we focus on $DB^{-3/2}$ in the following discussions. The $DB^{-3/2}$ values of the sampled PWNe are listed in Table \ref{tab:pwn_fit_result}. The $\Gamma$ fitting results of 3C 58 and G21.5$-$0.9 are in the same order as that derived in \citet{TangC2012} with full synchrotron spectrum.

We further test the sensitivity of the effect of parameters on the $\Sigma$ profile (Figure \ref{fig:parameter_degeneracy} (e) and (f)). The $DB^{-3/2}$ value changes not only the shape but also the displacement (or interception) of model profiles. Compare to the $\Gamma$ profile, we notice that the shape of the $\Sigma$ profile becomes insensitive to $DB^{-3/2}$ as long as $DB^{-3/2}$ is large enough. From Figures \ref{fig:parameter_degeneracy} (c), the offset of $\Gamma$, which implies the general hardness of the PWN, is dominated by  $\alpha$ of the injected high-energy particles. On the other hand, the displacement of the $\Sigma$ profile could be affected by $DB^{-3/2}$, $\alpha$, and $\eta$. Finally, $\eta$ only affects the displacement in the $\Sigma$ profile because it represents the energy transferring efficiency from the spin-down power of the pulsar to the PWN. 

\begin{figure*}
\centering
\includegraphics[width=0.6\textwidth]{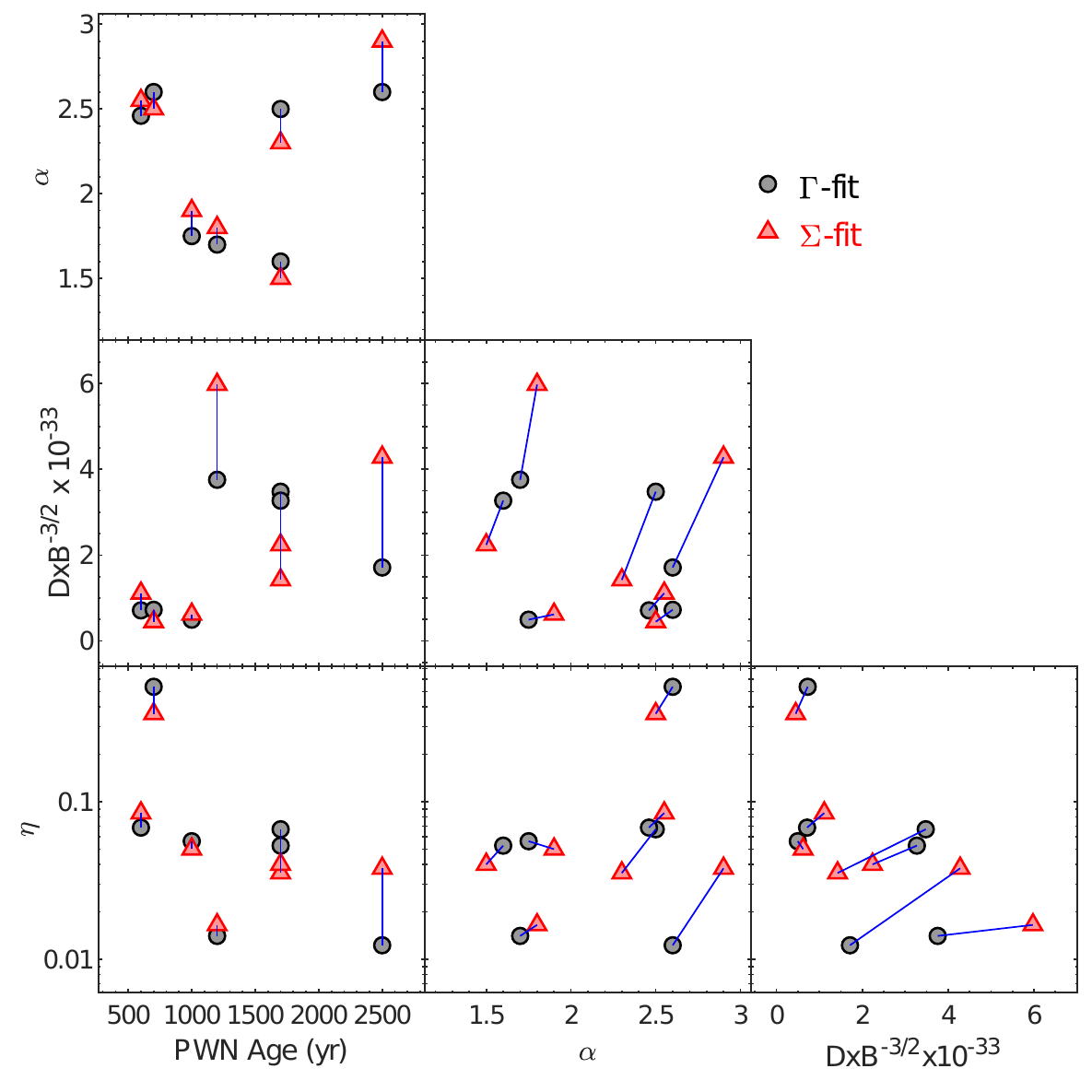} 
\caption{Correlation between model parameters and physical parameters. Gray circles denote the parameters derived from the $\Gamma$ fit, while red triangles are parameters from the $\Sigma$ fit. We use blue line to link the parameters obtained by \LD{$\Gamma$ and $\Sigma$ fits} of each PWN.  The ages of PWNe and $L_{sd}$ are adopted from Table \ref{tab:pwn_list}.  \label{fig:corr_all}}
\end{figure*}

\subsection{Profile Characteristics and Model Parameters} \label{apparent_vs_model}
Based on the discussion in Section \ref{parameter_degen}, we test whether the characteristics, i.e. slope and mean value of the $\Gamma$ and $\Sigma$ profiles, can be interpreted with the pure diffusion parameters. To achieve this goal, we simply fit the $\Gamma$ profile of each PWN with a straight line. Crab and Vela are not included in this analysis since their profiles cannot be described with our model. The slope can roughly represent the $\Gamma$ gradient. We take the mean of $\Gamma$ as the value at the half of the PWN radius predicted from the best-fit straight line. For the $\Sigma$ profile, we define the same slope and the mean $\Sigma$ as we did for the $\Gamma$ profile. The difference is that we use the logarithmic scale for the surface brightness.

Then, we plot the apparent features (slope and mean value of the $\Gamma$ and the $\Sigma$ profiles) versus the model parameters in Figure \ref{fig:apparent_vs_model}. The most significant correlation is, not surprisingly, between the mean of $\Gamma$ and $\alpha$. The linear correlation coefficient is $\rho=0.89$ with a null hypothesis probability $p=0.006$ for \LD{$\Gamma$ fit}, where the \LD{$\Sigma$-fitting} result is $\rho=0.96$ with $p=6\times10^{-4}$.  A relatively weak correlation can be seen between $DB^{-3/2}$ and the slope of the $\Gamma$ profiles with $\rho=-0.85$ and $p=0.02$ for the \LD{$\Gamma$ fit}. The correlation is insignificant for the \LD{$\Sigma$ fit} with $\rho=-0.40$ and $p=0.3$. This is caused by the large discrepancy between the \LD{$\Gamma$} and \LD{$\Sigma$ fitting results} of a few PWNe where the \LD{$\Sigma$ fit} cannot well describe the $\Gamma$ profile.

The slope of the $\Sigma$ profile is correlated with $DB^{-3/2}$, where the correlation analysis suggests $\rho=0.75$ and $p=0.05$ for both the \LD{$\Sigma$- and $\Gamma$- fits}. The relation between these two parameters might be non-linear. Therefore, we also calculate Kendall's rank correlation and obtain a Kendall's $\tau=0.81$ with $p=0.01$ for the \LD{$\Gamma$ fit}, and $\tau=0.52$ with $p=0.1$ for the \LD{$\Sigma$-fit}. This suggests a possible correlation and consistent with the result in Section \ref{parameter_degen}, in which we show that a larger $D$ (or smaller $B$) result in a flatter $\Sigma$ profile.  Interestingly, the slope of the $\Sigma$ profile anti-correlates with $\eta$ with $\rho=-0.84$ and $p=0.01$ for the \LD{$\Sigma$ fit} ($\rho=-0.88$ and $p=0.009$ for the \LD{$\Gamma$ fit}), suggesting a similar significance as that between $DB^{-3/2}$ and the $\Sigma$ profile. This is not expected as we discussed in Section \ref{parameter_degen} and could indicate physical connection underneath. Finally, the mean surface brightness likely anti-correlates with $DB^{-3/2}$ and positively correlates with $\eta$, these are expected in Section \ref{parameter_degen}. However, the mean surface brightness does not correlate with $\alpha$, suggesting that $\alpha$ is not the dominant factor for the $\Sigma$ prodile. 

\subsection{Correlation between Model Parameters and Physical Parameters}
We tested whether the best-fit parameters and physical parameters correlate with each other. We performed correlation analysis between four parameters, including $DB^{-3/2}$, $\alpha$, $\eta$, and the PWN age. The results are shown in Figure \ref{fig:corr_all}.   It is worth mentioning that $\eta$ of Kes 75 is the largest among selected PWNe. Given that PSR J1846$-$0258 is a high $B$-field pulsar that shows magnetar-like burst and outburst behavior, it is possible that the dissipation of the extremely strong magnetic field results in additional energy output from the pulsar \citep{GavriilGG2008, NgSG2008, KumarS2008}.

In general, we found no significant correlations between these parameters, indicating that these parameters are independent of each other. We found that $DB^{-3/2}$ may weakly anti-correlate with $\log(\eta)$ with $\rho=-0.7$ and $p=0.07$ for the \LD{$\Gamma$ fit}. This possible correlation is not observed for the \LD{$\Sigma$ fit} with $\rho=-0.5$ and $p=0.3$, suggesting no significance. We could not draw a strong conclusion with our current sample. However, this could be connected to the anti-correlation between the slope of the brightness profile and $\eta$ in Section \ref{apparent_vs_model}. A larger $DB^{-3/2}$ implies a steeper $\Gamma$ gradient in the radial profile, i.e.~stronger synchrotron energy loss. Therefore, a high energy conversion rate might not be needed.

\section{Summary}\label{summary}
We applied a pure diffusion model to fit the $\Gamma$ and $\Sigma$ profiles of nine young PWNe, including well studied 3C 58 and G21.5$-$0.9, a few faint sources G54.1+0.3, G291.0$-$0.1, G310.6$-$1.6, Kes 75, and three bright PWNe Crab, MSH 15$-$5\emph{2}, and Vela. We analyzed archival \chandra\ observations and fit the spectra of each PWN with a power-law model. Except for Crab and Vela, all the sampled PWNe show similar $\Gamma$ and $\Sigma$ profiles. $\Gamma$ increases slowly with the distance from the central pulsar. This behavior cannot be described by the MHD model that suggests an abrupt increase in $\Gamma$ \citep{SlaneHS2004, TangC2012, IshizakiTA2017} and therefore suggests that particle diffusion dominates the particle transfer rather than a steady flow for most PWNe. We derived the brightness profile from the pure diffusion model and found that the observed profiles are consistent with the model prediction. However, the parameters that can describe the $\Gamma$ profile and those for the $\Sigma$ profiles are not consistent with each other in a few cases like 3C 58 and G54.1+0.3. The discrepancy implies a simple diffusion model may not be enough to produce the entire distribution of PWNe and future improvement is necessary. The discrepancy could be attributed to the intrinsic spherical asymmetry, the particle reflection by the PWN boundaries, energy-dependent diffusion, and spatially-dependent $B$ field. The latter two could be specified with multi-wavelength studies. A model with a spatially varying magnetic field could be constructed in the future. We further investigate the relationship between the model parameters and the apparent feature of both profiles. The effect of the diffusion coefficient and the magnetic field are coupled with each other, and the observed $\Gamma$ and the $\Sigma$ gradient highly depend on the $DB^{-3/2}$ value. The mean hardness of a PWN significantly correlates with the distribution of the injected particles, while the brightness of a PWN is dominated by the energy conversion efficiency $\eta$. Finally, we found no significant correlations between model parameters and the physical quantities of PWNe.

\begin{acknowledgments}
\LD{We thank the anonymous reviewer for valuable comments that improved this paper}. This research is based on the data obtained from the Chandra Data Archive and has made use of the software provided by the \emph{Chandra} X-ray Center (CXC) in the application packages CIAO, ChIPS, and Sherpa. C.-P.H.~acknowledges support from the Ministry of Science and Technology in Taiwan through grant MOST 109-2112-M-018-009-MY3. W.I. is supported by JSPS KAKENHI Grants No.~21J01450. C.-Y.N.~is supported by a GRF grant of the Hong Kong Government under HKU 17305416. S.J.T.~ acknowledges support by the Aoyama Gakuin University-Supported Program ``Early Eagle Program''.
\end{acknowledgments}
\facilities{\emph{CXO} (ACIS)}
\software{CIAO \citep{FruscioneMA2006}, Sherpa \citep{FreemanDS2001}}

\bibliographystyle{aasjournal}

\clearpage

\appendix
\setcounter{table}{0}
\setcounter{figure}{0}
\renewcommand{\thetable}{A\arabic{table}}
\renewcommand\thefigure{A\arabic{figure}}  

\section{Data Sets Used in This Research}
We list all data sets used in this analysis in Table \ref{tab:chandra_log}.
\startlongtable
\begin{deluxetable}{llcc}
    \tablecaption{\chandra\ observation log of selected PWNe. \label{tab:chandra_log}} 
    \tablehead{
    \colhead{PWN Name} & 
    \colhead{ObsID} & 
    \colhead{Date} & 
    \colhead{Exposure (ks)} 
    }
\startdata
3C 58 & \dataset[728]{http://cda.harvard.edu/chaser/viewerContents.do?obsid=728} & 2000-09-14 & 38.7\\
 & \dataset[4383]{http://cda.harvard.edu/chaser/viewerContents.do?obsid=4383} & 2003-04-22 & 38.7 \\
 & \dataset[4382]{http://cda.harvard.edu/chaser/viewerContents.do?obsid=4382} & 2003-04-23 & 167.4 \\
 & \dataset[3832]{http://cda.harvard.edu/chaser/viewerContents.do?obsid=3832} & 2003-04-26 & 135.8 \\
\hline 
G21.5$-$0.9 & \dataset[159]{http://cda.harvard.edu/chaser/viewerContents.do?obsid=159} & 1999-08-23 & 14.9 \\
& \dataset[1230]{http://cda.harvard.edu/chaser/viewerContents.do?obsid=1230} & 1999-08-23 & 14.6 \\
& \dataset[1433]{http://cda.harvard.edu/chaser/viewerContents.do?obsid=1433} & 1999-11-15 & 15.0 \\
& \dataset[1716]{http://cda.harvard.edu/chaser/viewerContents.do?obsid=1716} & 2000-05-23 & 7.7 \\
& \dataset[1717]{http://cda.harvard.edu/chaser/viewerContents.do?obsid=1717} & 2000-05-23 & 7.5 \\
& \dataset[1718]{http://cda.harvard.edu/chaser/viewerContents.do?obsid=1718} & 2000-05-23 & 7.6 \\
& \dataset[1769]{http://cda.harvard.edu/chaser/viewerContents.do?obsid=1769} & 2000-07-05 & 7.4 \\
& \dataset[1770]{http://cda.harvard.edu/chaser/viewerContents.do?obsid=1770} & 2000-07-05 & 7.2 \\
& \dataset[1771]{http://cda.harvard.edu/chaser/viewerContents.do?obsid=1771} & 2000-07-05 & 7.2 \\
& \dataset[1838]{http://cda.harvard.edu/chaser/viewerContents.do?obsid=1838} & 2000-09-02 & 7.9 \\
& \dataset[1839]{http://cda.harvard.edu/chaser/viewerContents.do?obsid=1839} & 2000-09-02 & 7.7 \\
& \dataset[1553]{http://cda.harvard.edu/chaser/viewerContents.do?obsid=1553} & 2001-03-18 & 9.8 \\
& \dataset[1554]{http://cda.harvard.edu/chaser/viewerContents.do?obsid=1554} & 2001-07-21 & 9.1 \\
& \dataset[2873]{http://cda.harvard.edu/chaser/viewerContents.do?obsid=2873} & 2002-09-14 & 9.8 \\
& \dataset[3693]{http://cda.harvard.edu/chaser/viewerContents.do?obsid=3693} & 2003-05-16 & 9.8 \\
& \dataset[3700]{http://cda.harvard.edu/chaser/viewerContents.do?obsid=3700} & 2003-11-09 & 9.5 \\
& \dataset[5166]{http://cda.harvard.edu/chaser/viewerContents.do?obsid=5166} & 2004-03-17 & 10.0 \\
& \dataset[5159]{http://cda.harvard.edu/chaser/viewerContents.do?obsid=5159} & 2004-10-27 & 9.8 \\
& \dataset[6071]{http://cda.harvard.edu/chaser/viewerContents.do?obsid=6071} & 2005-02-26 & 9.6 \\
& \dataset[6741]{http://cda.harvard.edu/chaser/viewerContents.do?obsid=6741} & 2006-02-22 & 9.8 \\
& \dataset[8372]{http://cda.harvard.edu/chaser/viewerContents.do?obsid=8372} & 2007-05-25 & 10.0 \\
& \dataset[10646]{http://cda.harvard.edu/chaser/viewerContents.do?obsid=10646} & 2009-05-29 & 9.5 \\
& \dataset[14263]{http://cda.harvard.edu/chaser/viewerContents.do?obsid=14263} & 2012-08-08 & 9.6 \\
& \dataset[16420]{http://cda.harvard.edu/chaser/viewerContents.do?obsid=16420} & 2014-05-07 & 9.6\\
\hline 
G54.1+0.3 & \dataset[1983]{http://cda.harvard.edu/chaser/viewerContents.do?obsid=1983} & 2001-06-06 & 38.5 \\
 & \dataset[9886]{http://cda.harvard.edu/chaser/viewerContents.do?obsid=9886} & 2008-07-08 & 65.3 \\
 & \dataset[9108]{http://cda.harvard.edu/chaser/viewerContents.do?obsid=9108} & 2008-07-10 & 34.7 \\
 & \dataset[9109]{http://cda.harvard.edu/chaser/viewerContents.do?obsid=9109} & 2008-07-12 & 162.3 \\
 & \dataset[9887]{http://cda.harvard.edu/chaser/viewerContents.do?obsid=9887} & 2008-07-15 & 24.8 \\
 \hline
G291.0$-$0.1 & \dataset[2782]{http://cda.harvard.edu/chaser/viewerContents.do?obsid=2782} & 2002-04-08 & 49.5\\
 & \dataset[16497]{http://cda.harvard.edu/chaser/viewerContents.do?obsid=16497} & 2013-10-28 & 37.6 \\
 & \dataset[14822]{http://cda.harvard.edu/chaser/viewerContents.do?obsid=14822} & 2013-11-03 & 51.5 \\
 & \dataset[14824]{http://cda.harvard.edu/chaser/viewerContents.do?obsid=14824} & 2013-11-04 & 74.1 \\
 & \dataset[14823]{http://cda.harvard.edu/chaser/viewerContents.do?obsid=14823} & 2013-11-10 & 67.0 \\
 & \dataset[16512]{http://cda.harvard.edu/chaser/viewerContents.do?obsid=16512} & 2013-11-13 & 63.4 \\
 & \dataset[16541]{http://cda.harvard.edu/chaser/viewerContents.do?obsid=16541} & 2014-01-21 & 29.2 \\
 & \dataset[16566]{http://cda.harvard.edu/chaser/viewerContents.do?obsid=16566} & 2014-02-15 & 34.6 \\
\hline
G310.6$-$1.6 & \dataset[9058]{http://cda.harvard.edu/chaser/viewerContents.do?obsid=9058} & 2008-06-29 & 5.1 \\
 & \dataset[12567]{http://cda.harvard.edu/chaser/viewerContents.do?obsid=12567} & 2010-11-17 & 52.8 \\
 & \dataset[17905]{http://cda.harvard.edu/chaser/viewerContents.do?obsid=17905} & 2016-11-25 & 13.9 \\
 & \dataset[19919]{http://cda.harvard.edu/chaser/viewerContents.do?obsid=19919} & 2016-11-29 & 79.7 \\
 & \dataset[19920]{http://cda.harvard.edu/chaser/viewerContents.do?obsid=19920} & 2016-12-05 & 39.46 \\
\hline
Kes 75 & \dataset[748]{http://cda.harvard.edu/chaser/viewerContents.do?obsid=748} & 2000-10-15 & 37.3 \\
 & \dataset[7337]{http://cda.harvard.edu/chaser/viewerContents.do?obsid=7337} & 2006-06-05 & 17.4 \\
 & \dataset[6686]{http://cda.harvard.edu/chaser/viewerContents.do?obsid=6686} & 2006-06-07 & 54.1 \\
 & \dataset[7338]{http://cda.harvard.edu/chaser/viewerContents.do?obsid=7338} & 2006-06-09 & 39.3 \\
 & \dataset[7339]{http://cda.harvard.edu/chaser/viewerContents.do?obsid=7339} & 2006-06-12 & 44.1 \\
 & \dataset[10938]{http://cda.harvard.edu/chaser/viewerContents.do?obsid=10938} & 2009-08-10 & 44.6 \\
 & \dataset[18030]{http://cda.harvard.edu/chaser/viewerContents.do?obsid=18030} & 2016-06-08 & 85.0 \\
 & \dataset[18866]{http://cda.harvard.edu/chaser/viewerContents.do?obsid=18866} & 2016-06-11 & 61.5 \\
\hline
MSH 15$-$5\textit{2} & \dataset[754]{http://cda.harvard.edu/chaser/viewerContents.do?obsid=754} & 2000-08-14 & 19.0 \\
 & \dataset[3834]{http://cda.harvard.edu/chaser/viewerContents.do?obsid=3834} & 2003-04-21 & 9.5 \\
 & \dataset[4384]{http://cda.harvard.edu/chaser/viewerContents.do?obsid=4384} & 2003-04-28 & 9.9 \\
 & \dataset[3833]{http://cda.harvard.edu/chaser/viewerContents.do?obsid=3833} & 2003-10-17 & 19.1 \\
 & \dataset[5334]{http://cda.harvard.edu/chaser/viewerContents.do?obsid=5334} & 2004-12-28 & 49.5 \\
 & \dataset[5338]{http://cda.harvard.edu/chaser/viewerContents.do?obsid=5335} & 2005-02-07 & 42.6 \\
 & \dataset[6116]{http://cda.harvard.edu/chaser/viewerContents.do?obsid=6116} & 2005-04-29 & 47.0 \\
 & \dataset[6117]{http://cda.harvard.edu/chaser/viewerContents.do?obsid=6117} & 2005-10-18 & 45.6 \\
\hline
Vela & \dataset[128]{http://cda.harvard.edu/chaser/viewerContents.do?obsid=128} & 2000-04-30 & 10.6 \\
 & \dataset[1987]{http://cda.harvard.edu/chaser/viewerContents.do?obsid=1987} & 2000-11-30 & 18.9 \\
 & \dataset[2813]{http://cda.harvard.edu/chaser/viewerContents.do?obsid=2813} & 2001-11-25 & 17.9 \\
 & \dataset[2814]{http://cda.harvard.edu/chaser/viewerContents.do?obsid=2814} & 2001-11-27 & 19.9 \\
 & \dataset[2815]{http://cda.harvard.edu/chaser/viewerContents.do?obsid=2814} & 2001-12-04 & 27.0 \\
 & \dataset[2816]{http://cda.harvard.edu/chaser/viewerContents.do?obsid=2816} & 2001-12-11 & 19.0 \\
 & \dataset[2817]{http://cda.harvard.edu/chaser/viewerContents.do?obsid=2817} & 2001-12-29 & 18.9 \\
 & \dataset[2818]{http://cda.harvard.edu/chaser/viewerContents.do?obsid=2818} & 2002-01-28 & 18.7 \\
 & \dataset[2819]{http://cda.harvard.edu/chaser/viewerContents.do?obsid=2819} & 2002-04-03 & 19.9 \\
 & \dataset[2820]{http://cda.harvard.edu/chaser/viewerContents.do?obsid=2820} & 2002-08-06 & 19.5 \\
 & \dataset[3861]{http://cda.harvard.edu/chaser/viewerContents.do?obsid=3861} & 2003-08-21 & 4.9 \\
 & \dataset[10132]{http://cda.harvard.edu/chaser/viewerContents.do?obsid=10132} & 2009-07-09 & 39.5 \\
 & \dataset[10133]{http://cda.harvard.edu/chaser/viewerContents.do?obsid=10133} & 2009-07-17 & 39.5 \\
 & \dataset[10134]{http://cda.harvard.edu/chaser/viewerContents.do?obsid=10134} & 2009-07-25 & 38.1 \\
 & \dataset[10135]{http://cda.harvard.edu/chaser/viewerContents.do?obsid=10135} & 2010-06-28 & 39.5 \\
 & \dataset[10136]{http://cda.harvard.edu/chaser/viewerContents.do?obsid=10136} & 2010-07-08 & 39.5 \\
 & \dataset[10137]{http://cda.harvard.edu/chaser/viewerContents.do?obsid=10137} & 2010-07-17 & 39.5 \\
 & \dataset[10138]{http://cda.harvard.edu/chaser/viewerContents.do?obsid=10138} & 2010-07-26 & 38.8 \\
 & \dataset[10139]{http://cda.harvard.edu/chaser/viewerContents.do?obsid=10139} & 2010-08-04 & 39.5 \\
 & \dataset[12073]{http://cda.harvard.edu/chaser/viewerContents.do?obsid=12073} & 2010-08-15 & 39.5 \\
 & \dataset[12074]{http://cda.harvard.edu/chaser/viewerContents.do?obsid=12074} & 2010-08-24 & 39.5 \\
 & \dataset[12075]{http://cda.harvard.edu/chaser/viewerContents.do?obsid=12075} & 2010-09-04 & 41.8\\
 \hline
Crab & \dataset[14458]{http://cda.harvard.edu/chaser/viewerContents.do?obsid=14458} & 2012-11-26 & 1.24 \\
 & \dataset[14679]{http://cda.harvard.edu/chaser/viewerContents.do?obsid=14679} &  	2013-01-02 & 0.66 \\
 & \dataset[14680]{http://cda.harvard.edu/chaser/viewerContents.do?obsid=14680} &  	2013-02-25 & 0.66 \\
 & \dataset[14685]{http://cda.harvard.edu/chaser/viewerContents.do?obsid=14685} &  	2013-03-07 & 0.61 \\
 & \dataset[14681]{http://cda.harvard.edu/chaser/viewerContents.do?obsid=14681} &  	2013-04-23 & 0.61 \\
 & \dataset[14682]{http://cda.harvard.edu/chaser/viewerContents.do?obsid=14682} &  	2013-08-16 & 0.61 \\
 & \dataset[14678]{http://cda.harvard.edu/chaser/viewerContents.do?obsid=14678} &  	2013-10-08 & 0.63 \\
 & \dataset[16245]{http://cda.harvard.edu/chaser/viewerContents.do?obsid=16245} &  	2013-10-19 & 1.29 \\
 & \dataset[16257]{http://cda.harvard.edu/chaser/viewerContents.do?obsid=16257} &  	2013-11-04 & 1.29 \\
 & \dataset[16357]{http://cda.harvard.edu/chaser/viewerContents.do?obsid=16357} &  	2014-01-29 & 1.3 \\
 & \dataset[16358]{http://cda.harvard.edu/chaser/viewerContents.do?obsid=16358} &  	2014-04-19 & 1.32 \\
 & \dataset[16359]{http://cda.harvard.edu/chaser/viewerContents.do?obsid=16359} &  	2014-08-16 & 1.32 \\
 & \dataset[16258]{http://cda.harvard.edu/chaser/viewerContents.do?obsid=16258} &  	2014-11-06 & 1.34 \\
 & \dataset[16360]{http://cda.harvard.edu/chaser/viewerContents.do?obsid=16360} &  	2015-01-18 & 1.37 \\
 & \dataset[16361]{http://cda.harvard.edu/chaser/viewerContents.do?obsid=16361} &  	2015-04-10 & 1.42 \\
 & \dataset[16362]{http://cda.harvard.edu/chaser/viewerContents.do?obsid=16362} &  	2015-08-60 & 1.44 \\
 & \dataset[16259]{http://cda.harvard.edu/chaser/viewerContents.do?obsid=16259} &  	2015-11-06 & 1.44 \\
 & \dataset[16363]{http://cda.harvard.edu/chaser/viewerContents.do?obsid=16363} &  	2016-01-31 & 1.44 \\
 & \dataset[16364]{http://cda.harvard.edu/chaser/viewerContents.do?obsid=16364} &  	2016-04-20 & 1.43 \\
 & \dataset[16365]{http://cda.harvard.edu/chaser/viewerContents.do?obsid=16365} &  	2016-08-14 & 1.51\\
 \hline
\enddata
\end{deluxetable}

\section{\emph{CHANDRA} images and spectral extraction regions}
\begin{figure*}[ht]
\begin{minipage}{0.33\linewidth}
\includegraphics[width=0.95\textwidth]{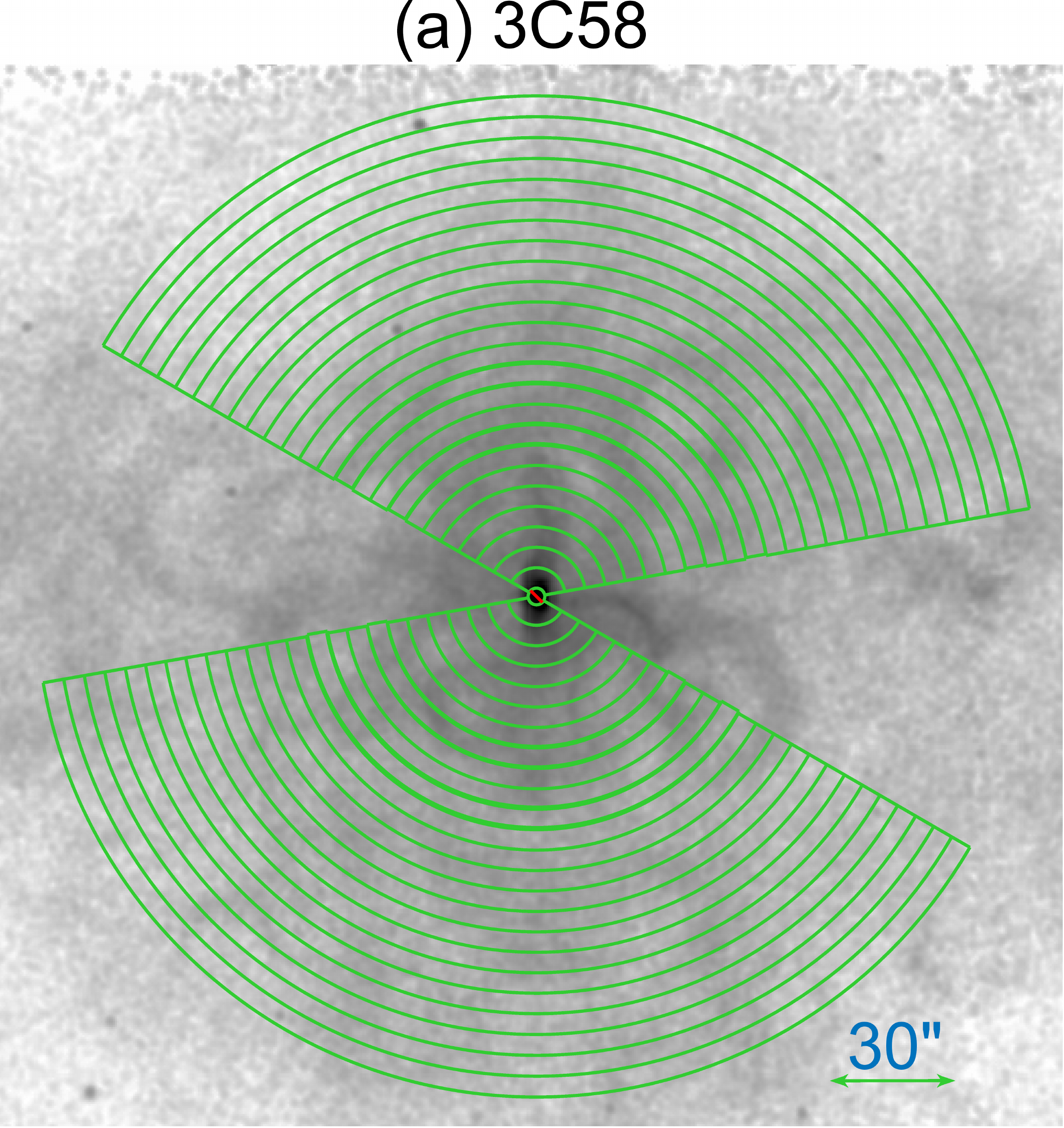} \\[0.2cm]
\includegraphics[width=0.95\textwidth]{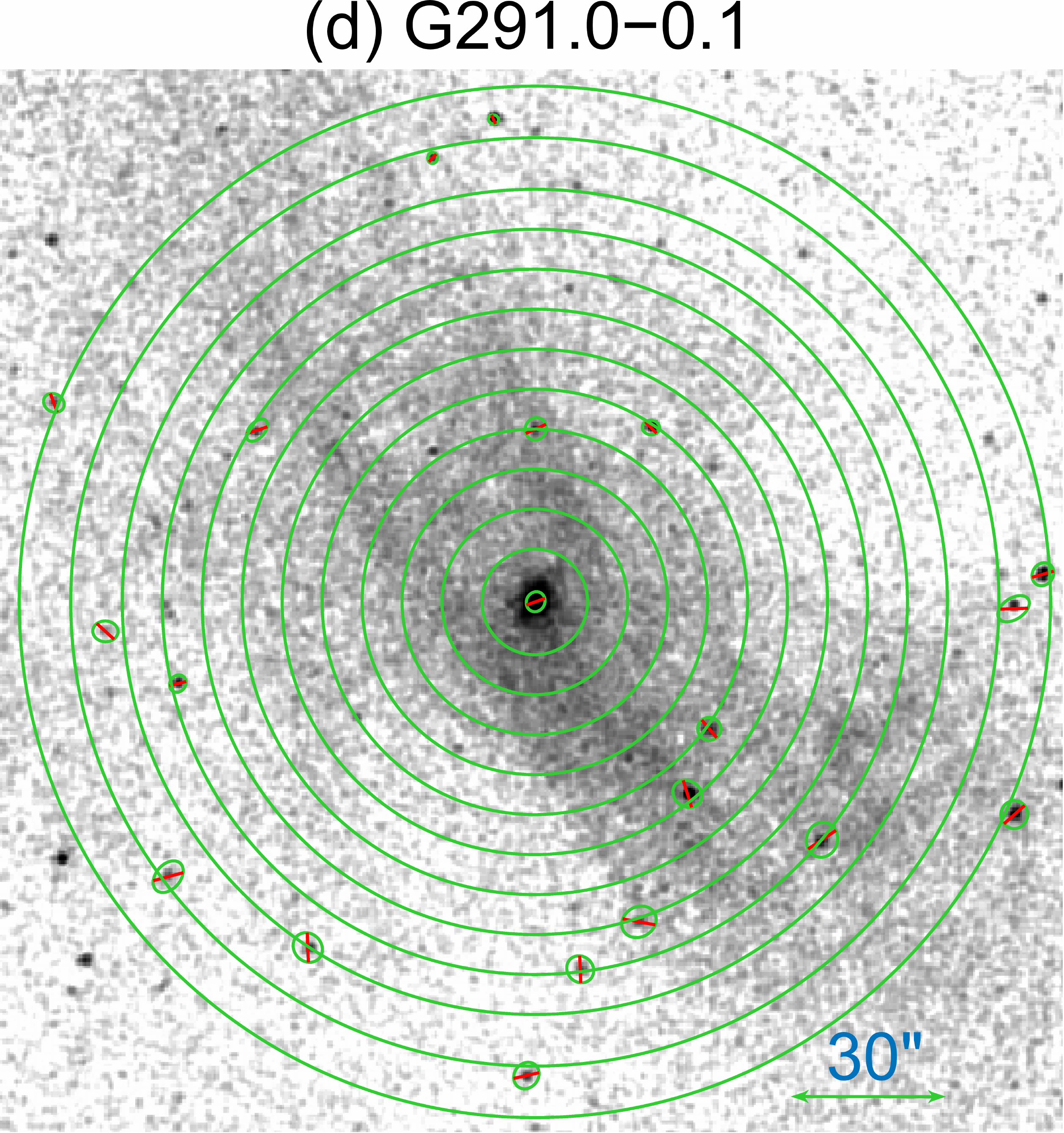} \\[0.2cm]
\includegraphics[width=0.95\textwidth]{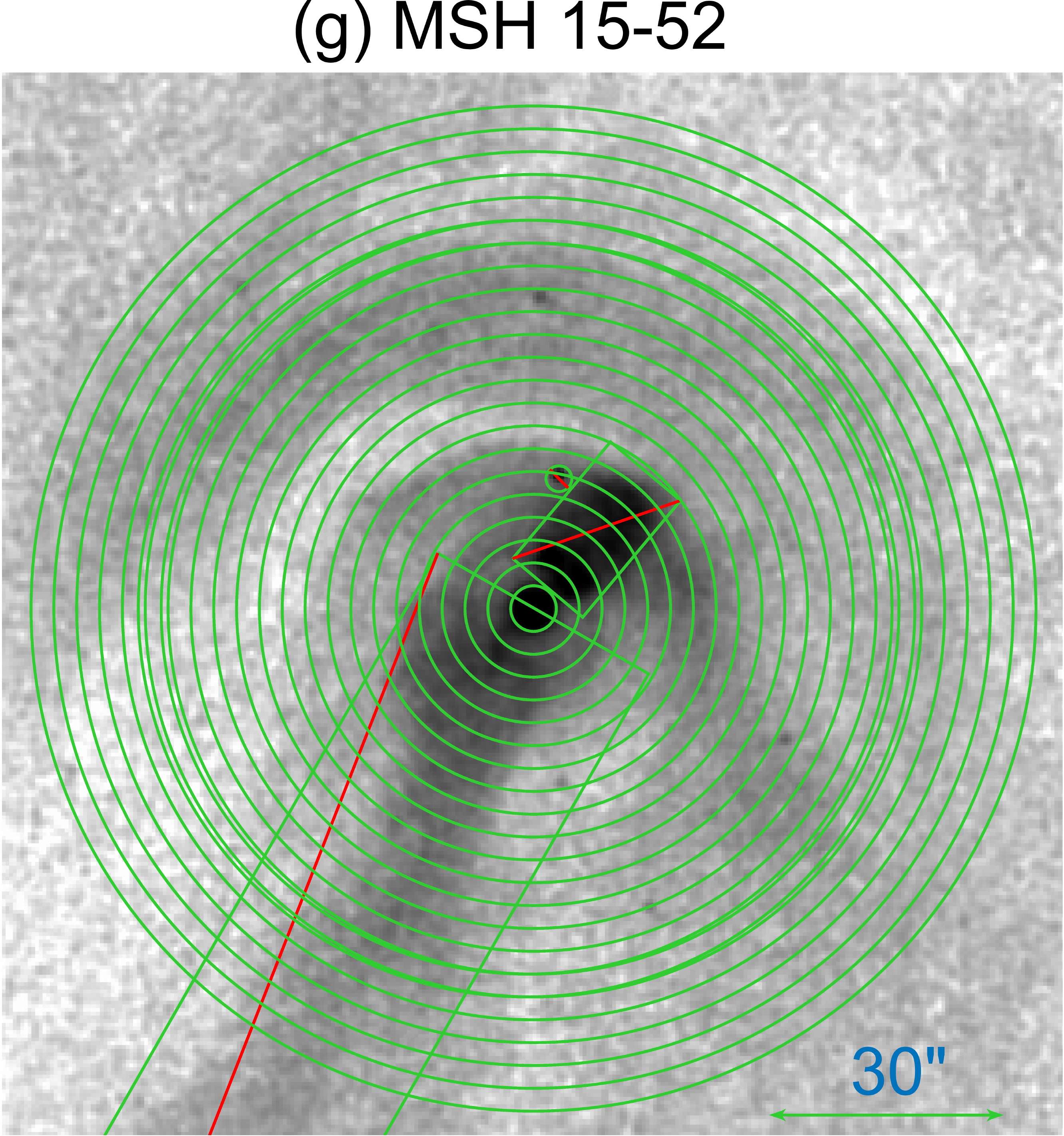} 
\end{minipage}
\smallskip
\begin{minipage}{0.33\linewidth}
\includegraphics[width=0.95\textwidth]{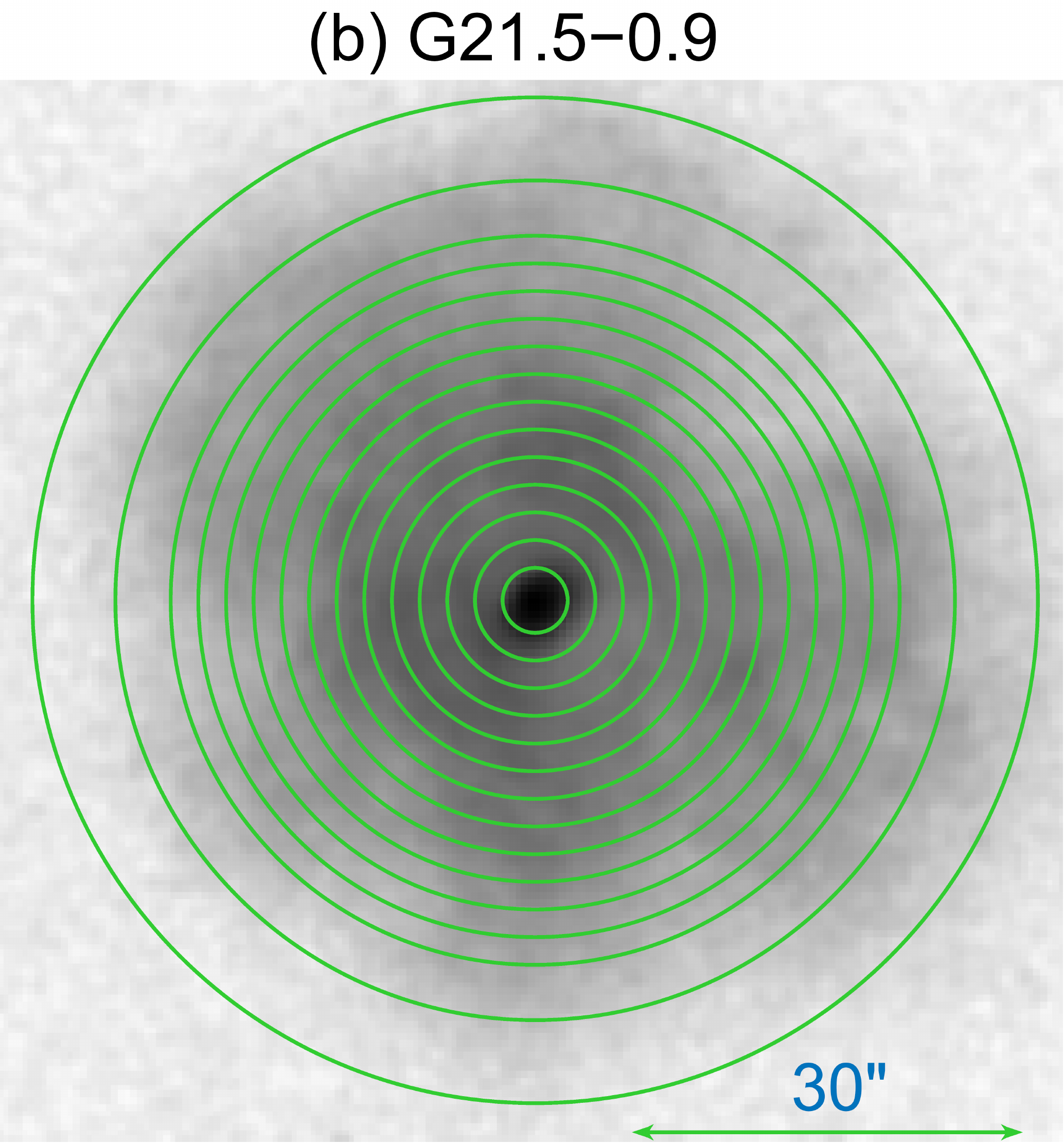} \\[0.2cm]
\includegraphics[width=0.95\textwidth]{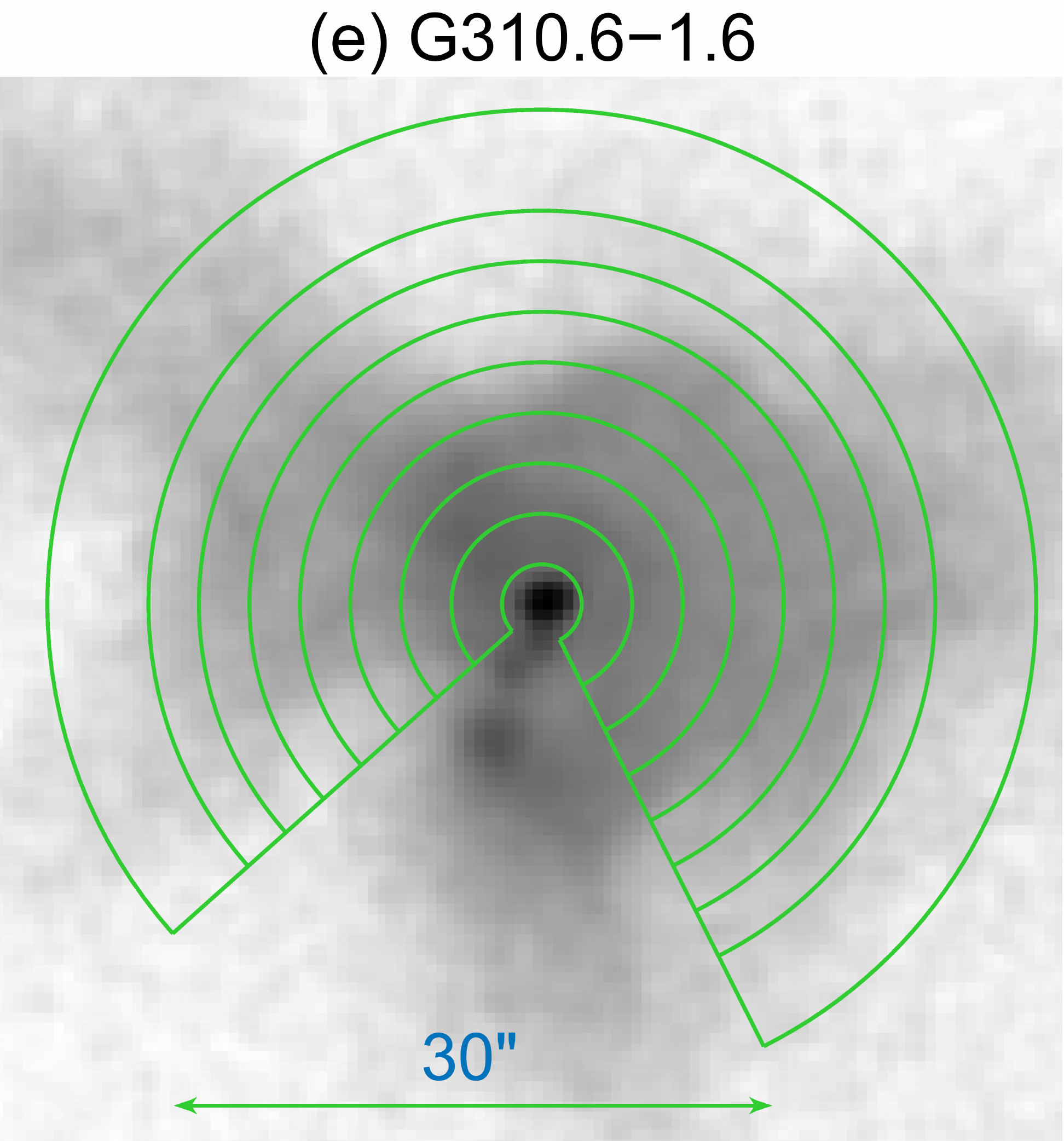} \\[0.2cm]
\includegraphics[width=0.95\textwidth]{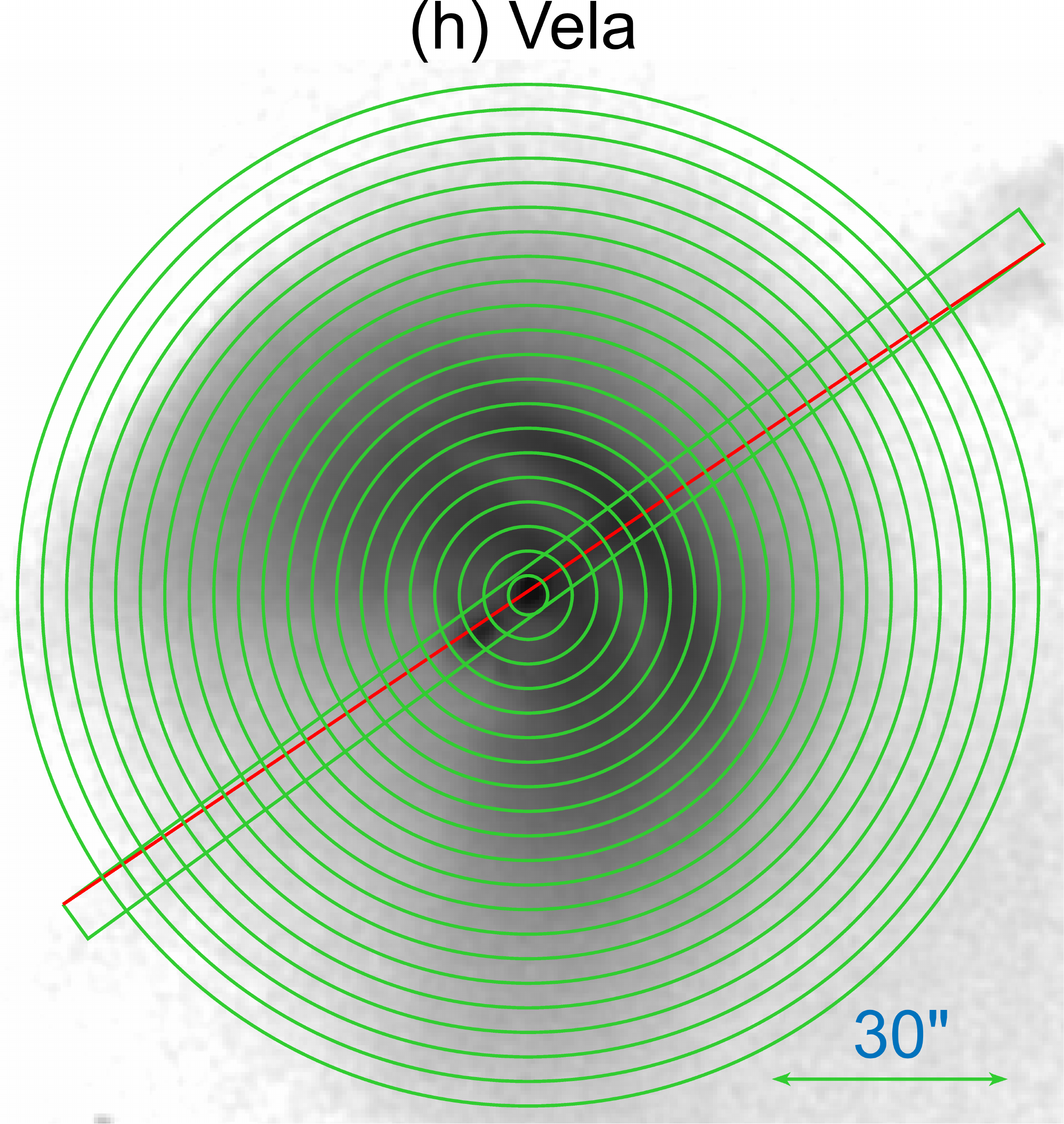} 
\end{minipage}
\smallskip
\begin{minipage}{0.33\linewidth}
\includegraphics[width=0.95\textwidth]{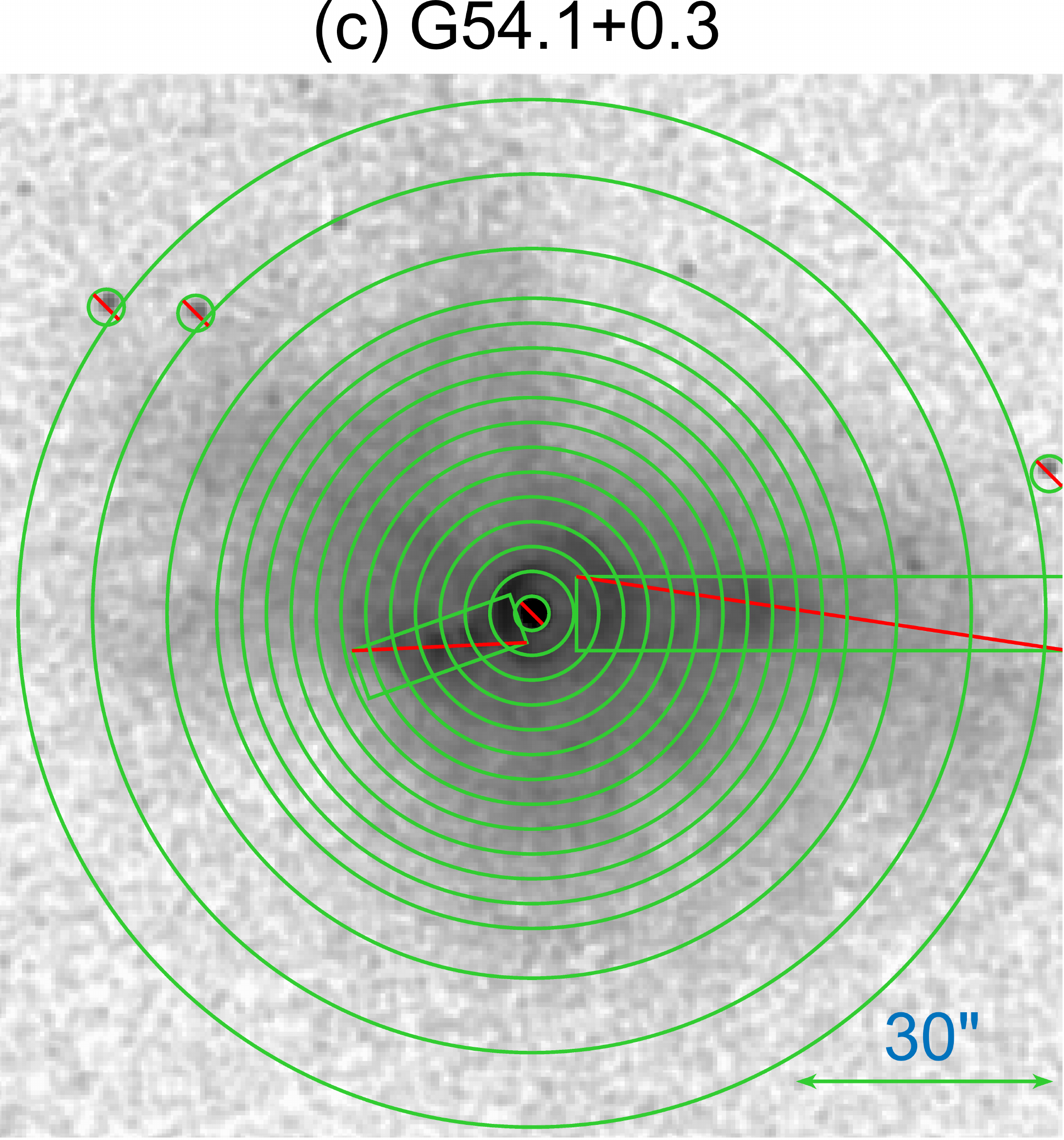} \\[0.2cm]
\includegraphics[width=0.95\textwidth]{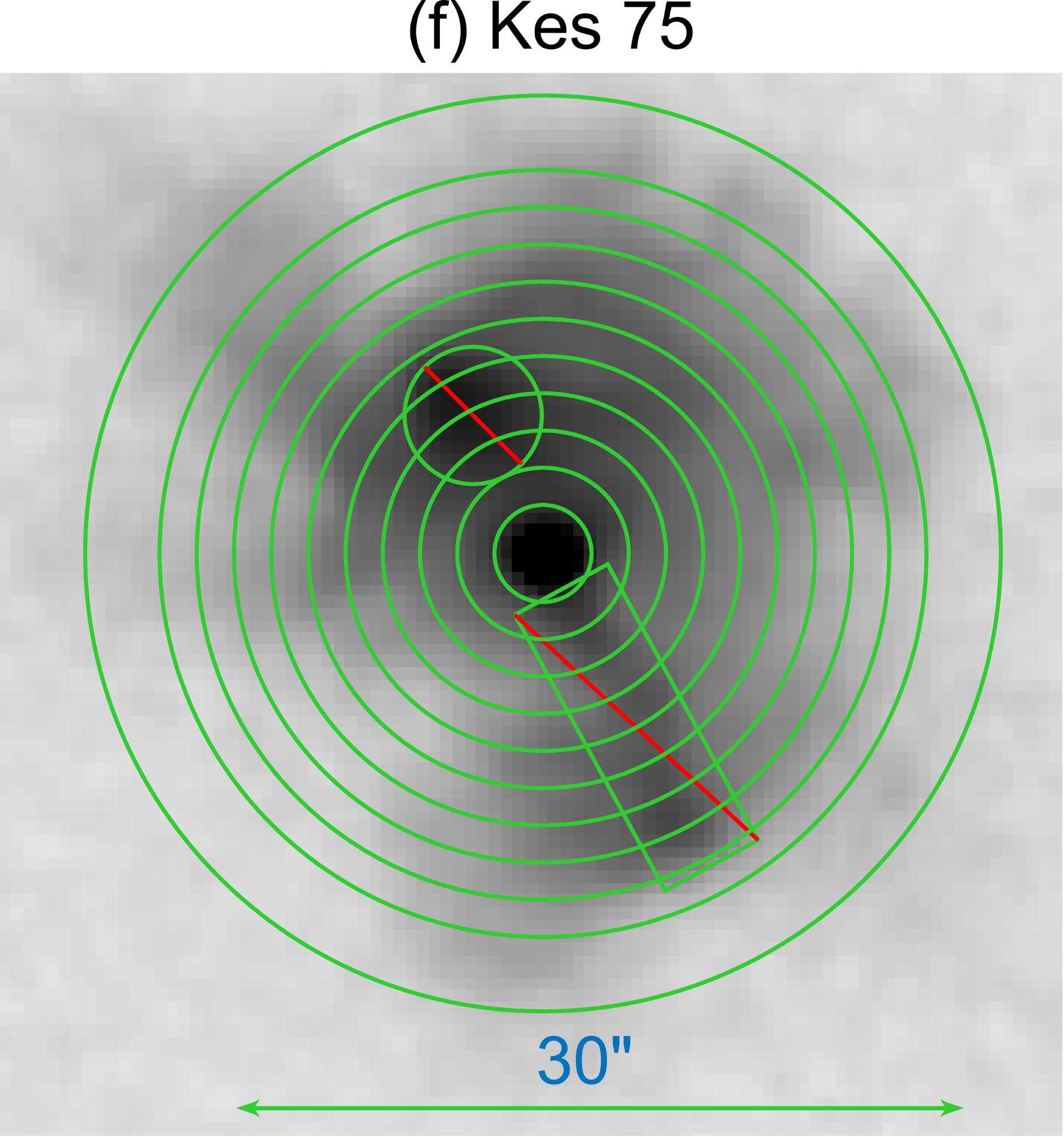} \\[0.2cm]
\includegraphics[width=0.95\textwidth]{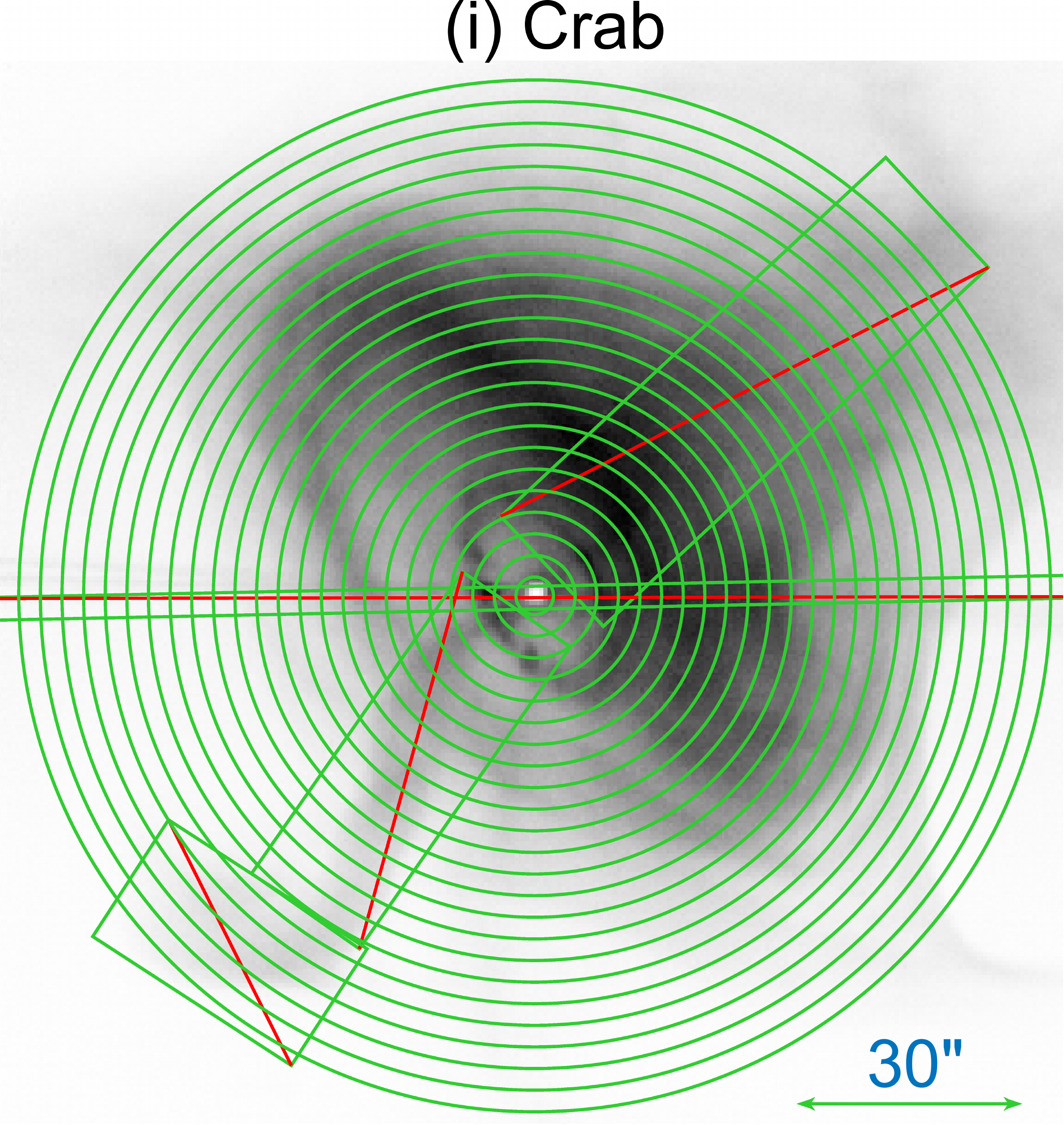}
\end{minipage}
\caption{\chandra\ 0.5--8 keV image of (a) 3C 58, (b) G21.5$-$0.9, (c) G54.1+0.3, (d) G291.0$-$0.1, (e) G310.6$-$1.6, (f) Kes 75, (g) MSH 15$-$5\emph{2}, (h) Vela, and (i) Crab. The green annulus denote the selected areas for spectral analysis. The jets, clumps, and background stars are removed. \label{fig:pwn_image}}
\end{figure*}

Figure \ref{fig:pwn_image} shows the \chandra\ 0.5--8 keV images of all the sampled PWNe. To study the radial profile, we use multiple annulus centered on the central pulsar to select X-ray photons and  collect X-ray spectra. 

\end{document}